\documentclass[prodmode,acmtoit]{acmsmall} 

\usepackage{mathdefs}
\usepackage{amsmath}
\usepackage{amssymb}
\usepackage{graphicx} 
\usepackage{algpseudocode}
\usepackage{algorithm}
\usepackage[noadjust]{cite}
\usepackage{xcolor}

\newcommand{\ie}{{\em i.e., }}
\newcommand{\cf}{{\em c.f., }}
\newcommand{\eg}{{\em e.g., }}

\newcommand{\al}{\underline{a}}
\newcommand{\au}{\bar{a}}
\newcommand{\tl}{\underline{t}}
\newcommand{\tu}{\bar{t}}
\newcommand{\xl}{\underline{x}}
\newcommand{\xu}{\bar{x}}
\newcommand{\dd}[1]{\frac{\partial}{\partial #1}}
\newcommand{\ddt}[1]{\frac{\partial^2}{\partial #1^2}}
\newcommand{\ddtt}[2]{\frac{\partial^2}{\partial #1 \partial #2}}
\newcommand{\rhol}{\underline{\rho}}
\newcommand{\Vl}{\underline{V}}

\newtheorem{assumption}{Assumption}

\newcommand{\assref}[1]{Ass.~\ref{ass:#1}}

\newcommand{\corref}[1]{Cor.~\ref{cor:#1}}
\newcommand{\defref}[1]{Def.~\ref{def:#1}}
\newcommand{\exaref}[1]{Ex.~\ref{exa:#1}}

\newcommand{\prpref}[1]{Prop.~\ref{prp:#1}}

\newcommand{\secref}[1]{\S\ref{sec:#1}}
\newcommand{\appref}[1]{App.~\ref{app:#1}}
\newcommand{\figref}[1]{Fig.~\ref{fig:#1}}

\newcommand{\remref}[1]{Remark~\ref{rem:#1}}

\usepackage{hyperref}
\hypersetup{colorlinks=false}
\hypersetup{colorlinks,citecolor=black,filecolor=black,linkcolor=black,urlcolor=black}

\doi{0000001.0000001}

\issn{1234-56789}

\begin{document}

\markboth{S.\ Weber}{Facilitating adoption of Internet technologies and services with externalities via cost subsidization}

\title{Facilitating adoption of Internet technologies and services with externalities via cost subsidization}
\author{STEVEN WEBER
\affil{Drexel University}
}

\begin{abstract}
This paper models the temporal adoption dynamics of an abstracted Internet technology or service, where the instantaneous net value of the service perceived by each (current or potential) user / customer incorporates three key features: $i)$ user service affinity heterogeneity, $ii)$ a network externality, and $iii)$ a subscription cost.  Internet technologies and services with network externalities face a ``chicken and egg'' adoption problem in that the service {\em requires} an established customer base in order to {\em attract} new customers.  In this paper we study cost subsidization as a means to ``reach the knee'', at which point the externality drives rapid service adoption, and thereby change the equilibrium service fractional adoption level from an initial near-zero level to a final near-one level (full adoption).  We present three simple subsidy models and evaluate them under two natural performance metrics: $i)$ the duration required for the subsidized service to reach a given target adoption level, and $ii)$ the aggregate cost of the subsidy born by the service provide. First, we present a ``two target adoption subsidy'' (TTAS) that subsidizes the cost so as to keep the fraction of users with positive net utility at a (constant) target level until the actual adoption target is reached.  Second, we study a special case of the above where the target ensures all users have positive net utility, corresponding to a ``quickest adoption'' subsidy (QAS).  Third, we introduce an approximation of QAS, called AQAS, that only requires the service provider adjust the subsidy level a prescribed number of times.  Fourth, we study equilibria and their stability under uniformly and normally distributed user service affinities, highlighting the unstable equilibrium in each case as the natural target adoption level for the provider.  Finally, we provide a fictional case study to illustrate the application of the results in a (hopefully) realistic scenario, along with a brief discussion of the limitations of the model and analysis.
\end{abstract}



\acmformat{Steven Weber, 2016. Facilitating adoption of Internet technologies and services with externalities via cost subsidization.}

\begin{bottomstuff}
Author's address: S. Weber, Department of Electrical and Computer Engineering, Drexel University, Philadelphia, Pennsylvania 19104.
\end{bottomstuff}

\maketitle

\section{Introduction}
\label{sec:intro}
With the Internet fueling the rise of a ``network society''~\cite{Cas2010}, many Internet technologies and services\footnote{For conciseness we will use the term services to refer to both.} realize their value only after reaching a certain level of adoption.  In other words, they exhibit positive {\em externalities}, \eg Metcalfe's Law.  Externalities are well-known~\cite{KatSha1986,Cab1990} to affect service adoption, and in particular to create a ``chicken-and-egg'' problem (\ie a service requires customers in order to attract customers) that can often stymie the success of new services.  This is because, when a new service is offered, most potential adopters see a cost that exceeds its (low) initial value.  This barrier to entry has been used to explain the difficulties encountered by various Internet security protocols~\cite{OzmSch2006} as well as by new versions of the Internet itself, \ie IPv6~\cite{GueHos2010a}.  Understanding how to overcome this problem is an important challenge for any provider wishing to launch a new service with a network externality.  Towards this end, the Internet Architecture Board (IAB) held a workshop on Internet Technology Adoption and Transition (ITAT) in 2013 to \emph{``develop protocol deployment strategies that enable new features to rapidly gain a foothold and ultimately realize broad adoption.  Such strategies must be informed by both operational and economic factors.''}\footnote{\url{http://www.iab.org/activities/workshops/itat/}}  

In prior work \cite{iabitat2013,GueDeO2014} we investigated {\em service bundling} as a means of overcoming initial adoption inertia.  In this work, we analyze the service adoption dynamics (AD) under a standard diffusion model when the service provider employs {\em cost subsidization}.  The model used in both our prior work and in this paper has in common three key assumptions (see \secref{model} \assref{all}): $i)$ users are heterogeneous, \ie their affinity for the service varies, $ii)$ services exhibit positive externalities, \ie the utility perceived by a user is an increasing function of the service adoption level, and $iii)$ services have a subscription cost, \ie a user pays a fixed amount per unit time to participate in the service.  There are no additional costs to join the service, nor any contractual requirements that prevent leaving the service at any time.  The per-user cost is assumed to be non-discriminatory, \ie identical across users, and fixed (exogenous).  

This paper is an extension of prior work \cite{WebGue2014} on cost subsidization under diffusion AD.  That paper studied a {\em constant} subsidy under a {\em uniform} affinity distribution, and the main result was an expression for the subsidy duration and aggregate cost.  The key improvements of this paper relative to \cite{WebGue2014} include $i)$ generalization of the model to an {\em arbitrary} affinity distribution, $ii)$ introduction and analysis of three new subsidy models (TTAS, QAS, AQAS), $iii)$ analysis of the equilibria and associated stability for general affinity distributions, with thorough analysis of the uniform and normal case, $iv)$ a case study illustrating the applicability of the results, and $v)$ a brief study of the impact of nonlinear externalities.

\subsection{Related work}
\label{sec:related}

Subsidization is a natural solution for such services because it incentivizes adoption among initial adopters (``innovators'' \cite{Bas1969}), thereby allowing the adoption level to build up to the ``knee'', \ie the point at which the strength of the externality will incentivize the later adopters (``imitators''), and the subsidy will no longer be needed to sustain the service.  Subsidization may take many forms; we provide a (necessarily) selective and brief review of this large topic below.   

There is a long-standing awareness of the role of subsidies in realizing more efficient outcomes in ``markets'' that exhibit positive externalities \ie by demonstrating the benefits of Pigouvian subsidies \cite{Pig1920}.  For example,~\cite{ChaMit1998} examines the impact of early investments on a firm's growth rate in the telecommunication industry. It identifies that early investments can facilitate the creation of an initial user base, and lead to greater overall market share.  This awareness not withstanding, most of the focus to-date has been on case studies, \eg see~\cite{McISub2009} for a recent review.

There have been some recent efforts on the modeling front, stemming in part from interest in viral marketing in online (social) networks~\cite{CanBim2012,HarMir2008,SwaEry2012,AjoJad2014}.  These works are closely related to studies of adoption dynamics in social networks~\cite[Chapter 24]{Kle2007}, but with a focus on maximizing revenue rather than adoption.  The optimal marketing strategy in a symmetric network, \ie a product utility grows in proportion to its number of adopters, is investigated in~\cite{HarMir2008} by formulating it as the solution of a dynamic program. A general network setting is considered in~\cite{CanBim2012} with the important difference of considering a \emph{divisible} good, so that consumption maximization is the goal.  Although we consider a common cost to users, we note that multiclass (Paris Metro) pricing for heterogeneous users is addressed in~\cite{ChaWan2014}.  Finally, \cite{CouGya2016} considers pricing for users with heterogeneous affinities, which in their context takes the form of user and subscriber loyalty to Internet Service Providers (ISPs) and Content Service Providers (CSPs); this issue is also addressed in \cite{ChoQiu2016}.

Like \cite{Bas1969}, we focus on product adoption among heterogeneous users in the presence of an externality, but our work differs in that \cite{Bas1969} studies two classes with no adoption costs, and no subsidization.  Like \cite{KatSha1986}, we focus on subsidies (sponsorship in their paper) with externalities, but our work differs in that \cite{KatSha1986} looks at equilibrium pricing, whereas our interest is on adoption dynamics.  Like \cite{HarMir2008}, we address optimizing over subsidies, but \cite{HarMir2008} considers {\em buyer-specific} subsidies and externalities.  

Finally, we comment on the past and current role that subsidies have played in the adoption of Internet technologies and services.  A survey of network economics is given in \cite{Shy2011}.  One of the first and most influential articles establishing the connection between computer security and economics as a whole is \cite{And2001} (\cf \cite{AndMoo2007}); the author, Ross Anderson, founded the annual Workshop on the Economics of Information Security (WEIS) in 2001, dedicated to exploring this connection (\cf \cite{Sch2006}), and maintains the ``Economics and Security Resource Page'' \cite{AndEconSecWeb}.  As discussed in \cite{OzmSch2006}, the U.S. Department of Homeland Security subsidized open source software development for the Domain Name System Security Extensions (DNSSEC) as a means of facilitating its adoption.  A few years later, \cite{Cla2010} proposed government subsidize the cost of removing malware from end-user computers.  More recently, with the rise of social networks, subsidies are now a standard tool; the example of the smart phone application ride-sharing service Lyft is discussed in \cite{Ede2015}.  The subscription subsidy models used in startup dating websites are discussed in \cite{Wen2015}.   

\subsection{Contributions and outline}
\label{sec:contrib}

\S\ref{sec:model} introduces the basic mathematical model, with a justification of the model assumptions in \secref{assjust}, and an analysis of the special case of a service with no externality in \secref{noext}.  \prpref{noextAD} and \corref{noextUE} establish the uniqueness of the adoption equilibrium in this case.  Although it is not the only plausible scenario, it is natural for a service provider to consider using a subsidy in a situation with multiple stable equilibria, where the low initial service adoption level will result in unsubsidized adoption dynamics converging to the lower equilibrium, but a sufficiently strong subsidy may push the adoption level high enough to enable convergence to the higher equilibrium.  In contrast, multiple stable equilibria {\em are} possible with externalities; it is for this reason we believe it is most natural to study the use of subsidies for services exhibiting externalities, which we assume for the rest of the paper.  Our two figures of merit are the subsidy duration and the aggregate cost of the subsidy born by the provider (\defref{subcost}), \ie the instantaneous aggregate subsidy cost (the product of the subsidy per user times the number of users) integrated over the subsidy duration.  

\secref{two-target} introduces the two-target adoption subsidy (TTAS), where the subsidy is structured to maintain a constant fraction of users with positive net utility, until such time as the target adoption level is achieved.  We first establish a necessary condition for a subsidy to be extremal with respect to aggregate cost (\prpref{neccondext} in \secref{neccondext}), then show (\prpref{twotarsub} in \secref{ttas}) that the TTAS satisfies this condition, however there is no reason to suspect TTAS is optimal.  The aggregate cost of a TTAS (\prpref{twotargetaggcost}) shown in \figref{twotarcostdur} demonstrates the two key performance metrics, the aggregate cost and the subsidy duration, may or may not be in tension with each other, depending upon parameters.  

\secref{quickest} specializes the TTAS to the case when the subsidy is structured so that {\em all} users have positive net utility, which we term a quickest adoption subsidy (QAS).  We establish the nature of the aggregate cost of a QAS as a function of the target adoption level (\prpref{qasprop} in \secref{propQAS}) and identify parameter regimes where the subsidy cost is always positive, always negative, and admits a finite maximum, respectively (\figref{qascost}).  As TTAS (and thus QAS) impose the possibly unrealistic requirement that the provider instantaneously adjust the subsidy in response to the adoption level, we introduce the {\em approximate} QAS (AQAS) in \secref{approxtwotar}, wherein the provider sets a sequence of intermediate adoption levels and adjusts the subsidy amount in a piecewise constant manner.  We give the aggregate cost of AQAS (\prpref{approxQAScost}) as a function of these target parameters, and study the cost of AQAS over QAS in \figref{qad}.

\secref{aff}. Until now the target adoption level has been chosen exogenously, independent of the equilibria of the unsubsidized adoption dynamics.  The natural context for subsidization of services with externalities, however, is to use the subsidy to bootstrap the adoption of the service to reach a critical adoption level, at which point the externality is sufficiently strong to drive the adoption level to a high adoption level without subsidization.  Such a strategy, however, requires knowledge of the set of equilibria, their stability, and their dependence upon the three key model parameters: the affinity distribution, the nominal service cost, and the externality. \secref{aff} gives $i)$ the number of possible equilibria for a general affinity distribution (\prpref{numsolnconvexconcave} in \secref{genaff}), and $ii)$ a detailed investigation of the equilibria and adoption dynamics under uniformly (\secref{uniaff}) and normally (\secref{noraff}) distributed affinities, respectively.

\secref{case}.  A fictional case study is presented with the intention of illustrating the applicability of the preceding content in a plausible scenario of a mobile app / service startup.  First, the idealized continuous-time AD for an infinite population is shown (in \appref{proofs}) to connect with the more pragmatic discrete-time AD for a finite population.  Second, the startup holds a trial period with a certain structure in order to estimate key model parameters such as the externality and the user affinity distribution; it is shown that these quantities can in fact be estimated.  Third, the startup evaluates a suite of possible subsidies, namely constant subsidies and the three subsidies discussed above (TTAS, QAS, and AQAS), and simulation results demonstrate their performance.

\secref{limitation} offers a brief discussion of the limitations anticipated in applying the model and the analysis to real-world scenarios.

\secref{conc} holds a brief conclusion.

There are two appendices.  In \appref{nlext} we briefly investigate {\em nonlinear} externalities, since elsewhere we have assumed a {\em linear} externality per user, consistent with the quadratic sum-user utility growth of Metcalfe's law.  \appref{proofs} holds several longer proofs.

\section{Mathematical model}
\label{sec:model}

\subsection{Assumptions and justifications}
\label{sec:assjust}

Let $x(t) = x(t|\tl,\xl) \in [0,1]$ denote the fraction of the population that has adopted the service at each time $t \geq \tl$ subject to the initial condition $x(\tl) = \xl$.  The model captures AD in a large population of potential users of an Internet service exhibiting the assumptions in \S\ref{sec:intro}, formalized as \assref{all} below.    
\begin{assumption}
\label{ass:all}
\begin{enumerate}
\item The {\em net utility}, $V=V(x,u)$, perceived by a randomly selected user when the adoption level is $x$, the subsidized service cost is $c-u$, and the externality parameter is $e$, is the random variable $V(x,u) \equiv A - \nu(x,u)$, where $A$ is the random user affinity for the service, and $\nu(x,u) \equiv (c - u) - e x$ is the {\em user net cost}.
\item User service affinity {\em heterogeneity} is captured by the random variable $A$, with a continuous complementary cumulative distribution function (CCDF) $\bar{F}_A(a) \equiv \Pbb(A > a)$.  Affinities are independent and identically distributed (iid).  
\item The service {\em externality} is captured by the term $ex$, where $e \geq 0$ is the externality parameter.  We generalize this from (linear) $x$ to nonlinear $\kappa(x)$ in \appref{nlext}.  
\item The {\em subsidized cost} of adoption is $c-u$ where $c \geq 0$ is a constant representing the {\em nominal cost}, and $u \in \Rbb$ is the subsidy amount.  Observe $i)$ $u < 0$ corresponds to a {\em negative subsidy} (an increased service cost), where users pay more than the nominal cost, and $ii)$ $u > c$ corresponds to a {\em negative cost}, where the provider in fact pays users to join the service.  We write both $u(x)$ to represent a subsidy as a function of the state $x$ and $u(t)$ to represent a subsidy as a function of time $t$, with the intepretation clear from context.  Setting $u = 0$ corresponds to unsubsidized dynamics.
\item The adoption level follows standard {\em diffusion dynamics} \cite[Chapter 1, Equation 1]{MahPet1985}, with time-scale parameter $\gamma > 0$:
\begin{equation}
\label{eq:9fd9034f34g}
\dot{x}(t) = \gamma f(x(t),u(t)) \equiv \gamma ( \Pbb(V(x(t),u(t)) > 0) - x(t)) 
\end{equation} 
where the (positive or negative) net utility at adoption level $x$ and subsidy level $u$ is
\begin{equation}
\label{eq:adoptdyn}
f(x,u) \equiv \bar{F}_A(\nu(x,u)) - x.
\end{equation}
\end{enumerate}
\end{assumption}
\begin{remark}
\label{rem:modeljust}
Each of the five points in the above assumption are given a corresponding justification below.
\begin{enumerate}
\item Each of the three terms in $V(x,u)$ and $\nu(x,u)$, namely, $A$, $c-u$, and $ex$, reflects one of the key assumptions in \S\ref{sec:intro}. Linear utility models like $V(x,u)$ are standard in the network externality literature (\eg \cite{Bas1969,CanBim2012}), although more general models have been studied (\eg \cite{Cab1990}).  Note the net utility, and each of the three terms comprising it, represent values or costs per unit time.  
\item User affinity heterogeneity in the target population is central to the use of diffusion dynamics in \eqref{eq:adoptdyn}; a homogeneous population with fixed affinity $a$ results in the trivial case where the entire population has positive (negative) net utility for $x \gtrless (c-u-a)/e$, respectively.  The iid assumption is valid in many populations, but is also required for tractability.  
\item Metcalfe's ``Law'', which asserts the sum utility over all users of a network service grows in the square of the size of the user base, means the utility per user grows linearly, consistent with the $e x$ form of the externality in Ass.~\ref{ass:all}-3), where $e$ is the linear growth rate.  Metcalfe's Law is assumed in much, but not all, of the literature on adoption under externalities.  Other dependencies have been argued as more suitable \cite{BriOdl2006}; we study this (briefly) in \appref{nlext}.
\item The instantaneous subsidized subscription cost $c-u$ focuses our model on services where the provider charges each user a regular fee to participate (\eg World of Warcraft or Angie's List\footnote{www.angieslist.com, ~ www.warcraft.com}).  Although one can define an equivalent model with $A' = A - c$, we retain $c$ to facilitate investigation of cost subsidization $u$. 
\item The dynamics in \eqref{eq:adoptdyn} assert the rate of change of the adoption level is proportional to the difference between the fraction of the population that {\em would} adopt at adoption level $x(t)$, and the fraction of the population that {\em has} adopted, \ie $x(t)$.  This principle, admitting a wide class of variants, is standard in the literature, \eg \cite{Rog1962,Bas1969,Moo1991}.
\end{enumerate}
\end{remark}

Equilibria and stability are defined for {\em unsubsidized} AD, since they are asymptotic (in time) quantities, and we focus in this work on finite-duration subsidies.
\begin{definition}
\label{def:equil}
An adoption level $x \in [0,1]$ is an (unsubsidized) {\em equilibrium} if $\dot{x}(t) = 0$ when $u=0$, \ie $f(x,0) = 0$. The set of equilibria is denoted $\Xmc$.  An equilibrium $x$ is {\em stable} if $\frac{\drm}{\drm x} f(x,0) < 0$.  The stable equilibria set is $\bar{\Xmc} \subseteq \Xmc$.
\end{definition}
An equilibrium occurs when all users with positive (negative) net utility have (not) adopted the service.  

We presume the service provider has in mind a target adoption level, denoted $\xu$, with $\xu \in (\xl,1]$.  We consider the class of continuous controls, denoted $\Umc$, with each control $u \in \Umc$ representing a subsidy, for which the adoption level $x$ is driven to the target adoption level $\xu$ by some finite time $\tu[u] > \tl$, \ie $x(\tu[u]) = \xu$.  Following convention, we place the function argument of functionals in square brackets, as in $\tu[u]$.  Although it is clear from the definitions of $V(x,u)$ and $\nu(x,u)$ that $u$ depends essentially upon the state $x$, rather than on the time $t$, we nonetheless alternately denote $u = (u(t), t \in [\tl,\tu])$ as a function of time, and $u = (u(x), x \in [\xl,\xu])$ as a function of the state.  We emphasize that $\xu$ is exogenous, but not $\tu$.  We consider two metrics by which we assess the performance of the subsidy: the hitting time, $\tu[u]$, and the aggregate provider cost $J[u]$, both of which, naturally, are to be minimized.

\begin{definition}
\label{def:subcost}
The {\em aggregate cost of a subsidy} born by the provider for a subsidy function $u$ with induced adoption level $x$ over the duration $[\tl,\tu[u]]$ is  
\begin{equation}
\label{eq:subfuncost}
J[u] \equiv \int_{\tl}^{\tu} l(x(t),u(t)) \drm t
\end{equation}
for $l(x,u) \equiv x u$ 
the {\em instantaneous aggregate cost} to the provider when subsidy level $u$ is applied and the adoption level is $x$.  Changing variables from $t$ to $y$ via $\drm t = \drm x/\dot{x} = \drm x/(\gamma f(x,u))$, yields
\begin{equation}
\label{eq:subfuncostalt}
\gamma J[u] = \int_{\xl}^{\xu} \frac{l(x,u)}{f(x,u)} \drm x.
\end{equation} 
\end{definition}
Multiplying $J[u]$ by $\gamma$ as above makes the aggregate cost independent of $\gamma$, and we will often report results in this way. 

\subsection{Subsidies without externalities}
\label{sec:noext}

\begin{figure}[!ht]
\centering
\includegraphics[width=0.5\linewidth]{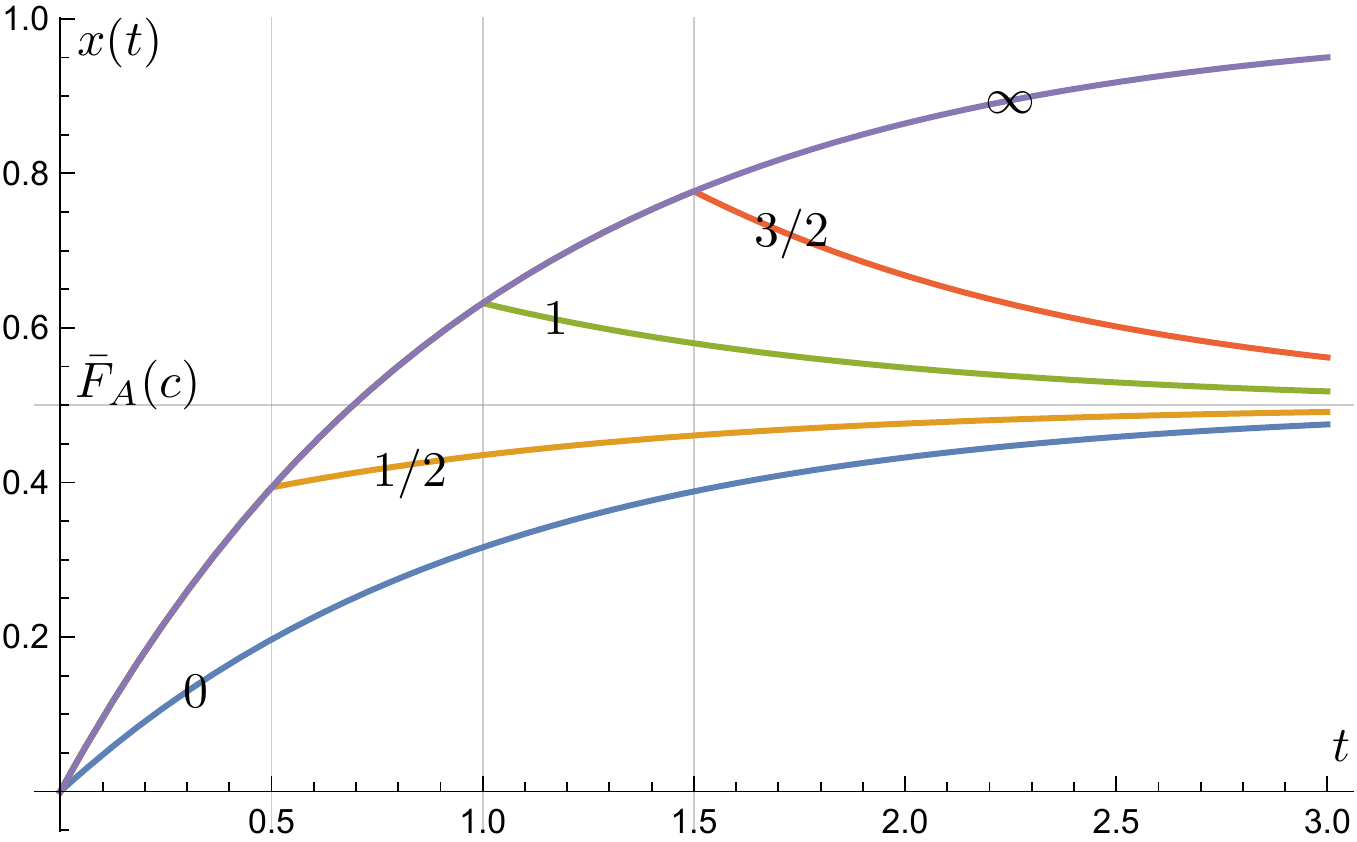}
\caption{\exaref{noext1} (\secref{noext}).  The AD $x(t)$ vs.\ $t$ for no externality ($e=0$), with $A \sim \mathrm{Uni}[0,1]$, and full subsidy with $u(t) = c = 1/2$ over $t \leq \tu$, with $\tu \in \{0,1/2,1,3/2,\infty\}$.  As soon as the subsidy ends, the adoption level begins converging to the sole equilibrium $\bar{F}_A(c) = 1/2$.}
\label{fig:noextuniaffconsub}
\end{figure}

In this section we study the impact of subsidization in the absence of an externality ($e=0$), so the subsidized utility is $V(x,u) = A - (c-u)$.  
\begin{proposition}
\label{prp:noextAD}
In the absence of an externality ($e=0$) and subsidization ($u=0$) \eqref{eq:adoptdyn} has solution
\begin{equation}
\label{eq:noextnosubadoptdyn}
x(t|\tl,\xl) = \bar{F}_A(c) - (\bar{F}_A(c) - \xl)\erm^{-\gamma(t-\tl)}, ~ t \geq \tl.
\end{equation}
The AD with subsidization $u = (u(t), t \in [\tl,\tu])$ that depends upon time $t$ and not on the adoption level $x$ (but is otherwise {\em arbitrary}) has solution
\begin{equation}
\label{eq:noextgensubsol}
x(t|\tl,\xl) =\xl \erm^{-\gamma (t-\tl)} + \gamma \erm^{-\gamma t}\int_{\tl}^t \erm^{\gamma \tau} \bar{F}_A(c-u(\tau)) \drm \tau,
\end{equation}
for $t \leq \tu$.  For $t > \tu$ the AD are $x(t) = x(t|\tu,\bar{\chi})$ for $x(t)$ in \eqref{eq:noextnosubadoptdyn} and $\bar{\chi} = x(\tu|\tl,\xl)$ the adoption level at the end of the subsidy at time $\tu$.
\end{proposition}
\begin{proof}
With no subsidization the AD \eqref{eq:adoptdyn} are $\dot{x}(t) = \gamma f(x(t),0)$, and the solution is obtained by separation of variables.  With the subsidy $u$ the AD are $\dot{x}(t) = \gamma f(x(t),u(t))$, and the solution is obtained by integrating factors. 
\end{proof}
The importance of this proposition is that it immediately yields the following corollary.
\begin{corollary}
\label{cor:noextUE}
In the absence of an externality ($e=0$) the unique equilibrium of both the unsubsidized and subsidized AD in \prpref{noextAD} is $\Xmc = \{\bar{F}_A(c)\}$.
\end{corollary}
\begin{example}
\label{exa:noext1}
To illustrate \prpref{noextAD} and Cor.\ \ref{cor:noextUE}, fix $\tl = \xl = e = 0$, $\gamma = 1$, and $c=1/2$, and suppose affinities are uniformly distributed, \ie $A \sim \mathrm{Uni}(0,1)$.  Fig.\ \ref{fig:noextuniaffconsub} shows the AD $x(t)$ vs.\ $t$ under a constant subsidy $u$ with $u(t) = c$ for $t \in [0,\tu]$ and $\tu \in \{0,1/2,1,3/2,\infty\}$.  After the subsidy ends, the AD converge to the equilibrium, $\bar{F}_A(c) = 1/2$.
\end{example}

\begin{remark}
\label{rem:noext}
In the absence of an externality, the resulting unique (stable) equilibrium $\bar{\Xmc} = \{\bar{F}_A(c)\}$ means that any finite-time subsidy cannot alter the final equilibrium level.  In the presence of an externality, however, the final equilibrium level {\em is} (in some cases) alterable by a finite-term subsidy (see \secref{uniaff} and \secref{noraff}).  In summary, subsidies are ineffective in the absence of an externality, and will be shown to be an effective and natural control mechanism when one is present.
\end{remark}

\begin{table}
\centering
\caption{Notation}
{\footnotesize 
\label{tab:notation}
\begin{tabular}{llllll} \hline
\secref{model} & $x(t|\tl,\xl)$ & adoption level at $t$ & \secref{two-target} & $\lambda = (\lambda(t))$  &  Lagrange multiplier fcn.\ \\ 
& $\tl,\xl$ & initial time \& adopt.\ level & & $\chi \in [\xu,1]$ & TTAS positive net utility frac.\  \\
& $\xu$ & target adoption level & & $\rho(\chi)$ & quantile under $\bar{F}_A$ \\
& $u = (u(x))$ & subsidy as fcn.\ of state & & $\xi(\chi)$ & normalized cost/ext.\ ratio \\
& $u = (u(t))$ & subsidy as fcn.\ of time & & $\tau(\chi)$ & normalized time $\tu[u]$ \\ 
& $e \geq 0$ & service externality param.\ & \secref{quickest} & $\rhol$ & minimum possible affinity \\
& $c \geq 0$ & nominal service cost & \eqref{eq:QASaggcost} & $\Jmc(\xu)$ & aggregate cost of the QAS \\
& $\nu(x,u)$ & user net cost & & $w$ & intermed.\ AQAS targets \\
& $A \sim \bar{F}_A$ & rand.\ user service affinity & & $\tilde{u}(x),\tilde{u}_i$ & intermed.\ AQAS subsidies  \\
& $V(x,u)$ & random user net utility & \eqref{eq:approxQAScost} & $\Jmc(w)$ & aggregate cost of the AQAS \\ 
\eqref{eq:adoptdyn} & $f(x,u)$ & diffusion term & \secref{aff} & $\Amc$ & support of $A$ \\
& $\gamma$ & time-scale param.\ for AD & & $(\Amc_i)$ & concave-convex partition of $\Amc$ \\
& $\Xmc$ & unsub.\ adoption equil.\ set & & $W$ & standard uniform rand.\ var.\ \\
& $\bar{\Xmc}$ & unsub.\ stable equil. set & \eqref{eq:xcircc} & $x^{\circ}$ & unstable equilibria \\
& $\tu[u]$ & subsidy end time & & $Z$ & standard normal rand.\ var.\ \\ 
& $l(x,u)$ & agg.\ inst.\ subsidy cost & \appref{nlext} & $\kappa(x)$ & nonlinear externality function \\ 
\eqref{eq:subfuncost} & $J[u]$ & aggregate subsidy cost & & & \\ \hline
\end{tabular}
}
\end{table} 

\section{Two-target adoption subsidies (TTAS)}
\label{sec:two-target}

In this section we first use Euler's equation (from the calculus of variations) to establish a necessary condition for a subsidy to be cost-extremal in \secref{neccondext}, then we introduce the class of two-target adoption subsidies, and establish them as extremal, in \secref{ttas}.

\subsection{A necessary condition for cost-extremal subsidies}
\label{sec:neccondext}

The following proposition gives a necessary condition for a subsidy to extremal\footnote{We emphasize that the condition is necessary to be {\em extremal}, \ie either minimal or maximal; additional ``second-order'' conditions are required to establish the functional to be minimal \cite{GelFom1963}.} with respect to $J[u]$ in \eqref{eq:subfuncost}.

\begin{proposition}
\label{prp:neccondext}
Suppose the affinity distribution $F_A$ is continuous and differentiable.  If the subsidy $u$ is extremal with respect to $J[u]$ then there exists a Lagrange multiplier function $\lambda = (\lambda(t), t \in [\tl,\tu])$ (where $\tu = \tu[u]$ satisfies $x(\tu[u]) = \xu$) such that, for each $t \in [\tl,\tu]$, 
\begin{equation}
\label{eq:neccondext}
u = e x - \gamma \lambda.
\end{equation}
\end{proposition}

\begin{proof}
We require only an elementary result in the calculus of variations.  In particular, we use \cite[Theorem 2]{GelFom1963} (c.f.\ the subsequent Remark 1) giving a necessary condition for $u$ to be extremal in the presence of ``finite subsidiary conditions'', also called non-holonomic constraints, \ie $g(\dot{x},x,u) = 0$, in our case given by $g(\dot{x},x,u) = \dot{x} - \gamma f(x,u)$.  In particular, the theorem states if $u$ is extremal for $J[u]$ under $g = 0$ then there exists a function $\lambda$ (defined above) such that $u$ is extremal for 
\begin{equation}
\tilde{J}[u] \equiv \int_{\tl}^{\tu} (l(x,u) + \lambda g(\dot{x},x,u)) \drm t.  
\end{equation}
The Euler equations for $\tilde{J}[u]$ are
\begin{equation}
\dd{u}l(x,u) + \lambda \dd{u}g(\dot{x},x,u) = 0, ~ \dd{x}l(x,u) + \lambda \dd{x}g(\dot{x},x,u) = 0. 
\end{equation}
Substitution of $\dd{u}l(x,u) = x$, $\dd{x}l(x,u) = u$, $\dd{u}g(\dot{x},x,u) = -\gamma f_A(\nu(x,u))$, and $\dd{x}g(\dot{x},x,u) = -\gamma (e f_A(\nu(x,u)) - 1)$, where $f_A(a) = \frac{\drm}{\drm a} F_A(a)$ is the user affinity PDF, into the Euler equations gives
\begin{eqnarray}
x = \gamma \lambda f_A(\nu(x,u)), ~ u = \gamma \lambda (e f_A(\nu(x,u)) - 1). 
\end{eqnarray}
Solving for $u$ gives \eqref{eq:neccondext}.
\end{proof}
The following subsection establishes a class of subsidies that satisfy this condition. 

\subsection{Two-target adoption subsidies}
\label{sec:ttas}

We propose a subsidy family, termed the {\em two-target adoption subsidy} (TTAS) family, with parameter $\chi \in (\xu,1]$, where the subsidy is structured such that the  fraction of users with positive net utility is kept constant at $\chi$, \ie $u = u(x)$ is such that $\bar{F}_A(\nu(x,u(x))) = \chi$.  It is clear from \eqref{eq:adoptdyn} that $\chi \geq \xu$ is necessary and sufficient to ensure $x$ reaches $\xu$. If the subsidy were to be continued indefinitely beyond its terminal time $\tu[u]$, then $x \to \chi$ as $t \to \infty$, so $\chi$ is also the asymptotic adoption level.
\begin{definition}
\label{def:twotargetadsub}
Given a target adoption level $\xu > \xl$, the {\em two-target adoption subsidy} (TTAS) with parameter $\chi \in (\xu,1]$ and corresponding quantile $\rho = \rho(\chi) \equiv \bar{F}_A^{-1}(\chi)$ has 
\begin{equation}
u = u(x) = c - \rho - e x,
\end{equation}
which obeys $\nu(x,u(x)) = \rho$ and $\bar{F}_A(\nu(x,u(x))) = \chi$, and thus, by \eqref{eq:adoptdyn}, $\dot{x} = \gamma (\chi - x)$.
\end{definition} 
The name {\em two-target} is apt because one may think of this subsidy as ``aiming'' for the ``inflated'' target $\chi > \xu$, but stopping once the actual target $\xu$ is achieved.  Observe the subsidy $u$ is linearly decreasing in the adoption level $x$, which accords with the intuition that larger subsidies are appropriate for lower adoption levels, and can be decreased, eliminated, and even made negative at higher adoption levels, where the externality strength offsets the higher cost without negatively affecting net utility.
\begin{remark}
\label{rem:twotarsubrng}
Recall \assref{all} Point 4. Observe $u < 0$, \ie the subsidy in fact increases the nominal cost from $c$ to $c - u > c$, when $x > (c-\rho)/e$.  As $x \in [\xl,\xu]$, it follows that these cost increases will occur at some point over the course of the subsidy if $\bar{x} > (c-\rho)/e$.  Next, observe $u > c$, \ie the subsidy in fact reduces the nominal cost to below $0$, meaning the provider in fact pays users to subscribe, when $x < -\rho/e$, which can only happen if $\rho < 0$.  These negative costs will occur over the course of the subsidy if $\xl < -\rho/e$.  The more typical scenario of a ``partial subsidy'', \ie $u \in [0,c]$ for each $t \in [\tl,\tu]$, will occur provided 
\begin{equation}
-\frac{\rho}{e} \leq \xl < \xu \leq \frac{c-\rho}{e}.
\end{equation}  
\end{remark}

\begin{proposition}
\label{prp:twotarsub}
The TTAS with parameter $\chi \in (\xu,1]$ satisfies the necessary condition from \prpref{neccondext} to be extremal for $J[u]$. The terminal time $\tu[u]$, the adoption level $x$, the subsidy $u$, and the Lagrange multiplier function $\lambda$ are (recall $\rho = \rho(\chi)$): 
\begin{eqnarray}
\gamma(\tu[u]-\tl) = \log \frac{\chi-\xl}{\chi-\xu} &, &  
x(t) = \chi - (\chi - \xl) \erm^{-\gamma(t-\tl)} \nonumber \\
u(t) = (c-\rho) - e x(t) &, & 
\gamma\lambda(t) = 2 e x(t) - (c - \rho)
\end{eqnarray}
\end{proposition}

We emphasize that, although the TTAS satisfies the {\em necessary} condition for extremality, there is no cause to believe the TTAS is itself optimal.

\begin{figure}[!ht]
\centering
\includegraphics[width=\linewidth]{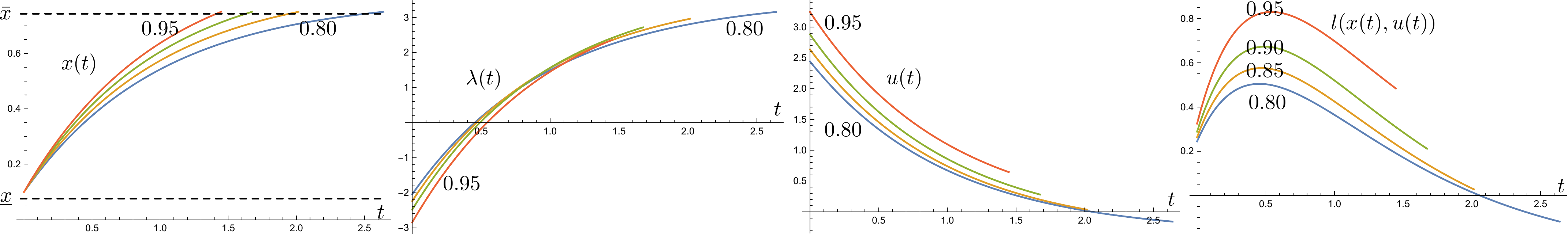}
\caption{\secref{ttas}: Illustration of the adoption dynamics under the two-target subsidy (\prpref{twotarsub}): $x(t)$ (far left), $\lambda(t)$ (middle left), $u(t)$ (middle right), and $l(x(t),u(t))$ (far right) vs.\ $t$, for TTAS adoption target $\chi \in \{0.80,0.85,0.90,0.95\}$.  Normal affinities ($A \sim N(0,1)$), with $\tl=0$, $\xl = 1/10$, $\xu=3/4$, $c=2$, $e=4$.}
\label{fig:twotarmdyn}
\end{figure}

\begin{proof}
The necessary condition $u = e x - \gamma \lambda$ from \prpref{neccondext} for a subsidy to be cost-extremal combined with the TTAS equation $u = c - \rho - ex$ from \defref{twotargetadsub} has solution
\begin{equation}
\label{eq:twotarnecprf}
2ex = c-\rho +\gamma \lambda, ~ 2u = c-\rho -\gamma \lambda.
\end{equation}
which highlights the symmetry between $(x,u)$.  As $\dot{x} = \gamma (\chi - x)$ under the TTAS, it follows that the dynamics are given by \prpref{noextAD}, but with $\chi$ replacing $\bar{F}_A(c)$.  That is, although \eqref{eq:noextnosubadoptdyn} holds for $e=0$ and $u=0$, it applies also in this case since the TTAS ensures $\bar{F}_A = \chi$.  The expression for $\tu[u]$ follows by solving $x(t) = \xu$, the expresion $u(t)$ follows from $u = c - \rho - ex$, and the expression for $\gamma \lambda(t)$ follows from $\gamma \lambda = 2 ex - (c-\rho)$. 
\end{proof}
The quantities $x,u,\lambda,l(x,u)$ are shown in \figref{twotarmdyn}.  Observe the inverse relationship between $x,u$ is such that the instantaneous cost $l(x,u)$ has a global maximum at some time $t \in (\tl,\tu)$.  In particular, for $t$ near $\tl$ the provider pays a large subsidy but only to a few users, for a moderate aggregate instantaneous cost, while for $t$ near $\tu$, the provider pays a very small subsidy to a much larger number of users, for a moderate aggregate instantaneous cost.  In fact, for $\chi$ small (here, $\chi = 4/5$), we see a negative instantaneous cost ($l < 0$) for $t > 2$, meaning the provider {\em raises} the cost above $c$ so as to extract revenue; the strength of the externality at a high adoption level offsets this high cost to keep the fraction of users with net positive utility at $\chi$.

\begin{figure}[!ht]
\centering
\includegraphics[width=\linewidth]{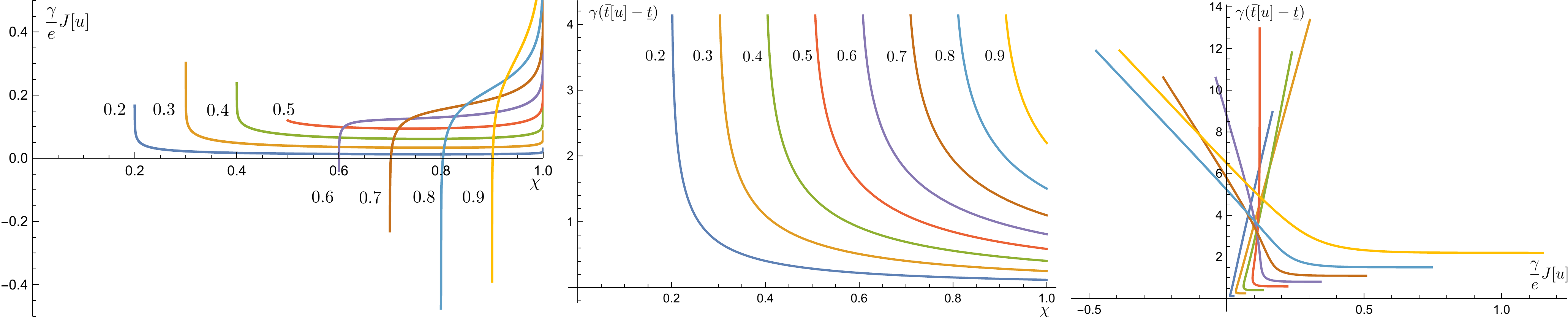}
\caption{\secref{ttas}: Illustration of the aggregate cost born by the provider (\prpref{twotargetaggcost}) and the subsidy duration (\prpref{twotarsub}) under the TTAS. Left: $\frac{\gamma}{e} J[u]$ vs.\ $\chi$ for various $\xu \in \{2,\ldots,9\}/10$.  Middle: $\gamma(\bar{t}[u]-\tl)$ vs.\ $\chi$ (same set $\xu$). Right: parametric plot of achievable $(J,\bar{t})$ pairs as $\chi$ is swept (same set $\xu$). Normal affinities ($A \sim N(0,1)$), with $\tl=0$, $\xl = 1/10$, $c=2$, and $e=4$.}
\label{fig:twotarcostdur}
\end{figure}

The TTAS has the additional benefit that its aggregate cost can be expressed succinctly, as given in the following result.
\begin{proposition}
\label{prp:twotargetaggcost}
The TTAS has aggregate cost
\begin{equation}
\label{eq:twotargetaggcost}
\frac{\gamma}{e} J[u] = (\xi(\chi) - \chi )( \chi \tau(\chi) - (\xu - \xl)) + \frac{1}{2}(\xu^2 - \xl^2) 
\end{equation}
where $\xi(\chi) \equiv (c-\rho(\chi))/e$ and $\tau(\chi) \equiv \log \frac{\chi-\xl}{\chi-\xu}$.
\end{proposition}
Observe $\tau(\chi) = \gamma (\bar{t}[u] - \tl)$ for $\bar{t}[u]$ the TTAS terminal subsidy time in \prpref{twotarsub}. 
\begin{proof}
Observe $l(x,u) = x (c - \rho - e x)$, and thus, by \eqref{eq:subfuncostalt} in \defref{subcost}, 
\begin{equation}
\label{eq:twotarcostprf1}
\frac{\gamma}{e} J[u] = \int_{\xl}^{\xu} \frac{x(\xi - x)}{\chi - x} \drm x,
\end{equation}
from which integration and algebra yields \eqref{eq:twotargetaggcost}.
\end{proof}
Plots of the aggregate cost $\frac{\gamma}{e} J[u]$ and normalized terminal time $\gamma(\tu[u]-\tl) = \tau(\chi)$ vs.\ $\chi$ for various $\xu$, with $\xl = 1/10$, $c = 2$, and $e=4$ are shown in \figref{twotarcostdur}.  Several points bear mention.  The qualitative behavior of $J[u]$ as a function of $\chi$ varies dramatically for various $\xu$, ranging from convex decreasing for small $\xu$, convex with internal minimum for moderate $\xu$, and increasing for larger $\xu$.  The corresponding $J$-optimal $\chi^*$ is therefore highly sensitive to $\xu$ even for fixed $(c,e)$.  Roughly speaking, $\chi^* = 1$ (quickest adoption) is $J$-optimal for small $\xu$, while $\chi^* = \xu$ (slowest adoption) is $J$-optimal for larger $\xu$.  In fact, $J$ may even be {\em negative}, \ie the provider earns a net {\em profit} from the subsidy, when $\xu$ is large and $\chi^*$ is sufficiently close to $\xu$.  The subsidy duration $\tu$ is naturally decreasing in $\chi$ for each $\xu$, and increasing in $\xu$.  The performance plot of achievable $(J,\tu)$ pairs, with $\chi$ as a parameter, shows that for small $\xu$ there is no tension between $J$ and $\tu$ (they can both be minimized by selecting a large $\chi$), while for large $\xu$ the two are in tension.  In particular, the provider may profit from the subsidy ($J < 0$) but at the expense of a larger adoption time $\tu$.

On account of the complicated nature of $J$ for the TTAS, and on account of its natural interest, we investigate in the next section the special case when $\chi = 1$.

\section{Quickest adoption subsidies}
\label{sec:quickest}

In this section we specialize the TTAS family to the special case of $\chi = 1$, which corresponds to minimizing $\tu[u]$, and which we therefore refer to as a {\em quickest adoption subsidy} (QAS).  As evident from \figref{twotarcostdur}, such a subsidy may or may not be optimal with regard to aggregate cost $J[u]$.  A QAS is natural from the perspective of the the service provider interested in achieving the target adoption level $\xu$ as soon as possible.  In this section we first establish some properties of aggregate cost of the QAS as a function of the target $\xu$ in \secref{propQAS}, then present an approximate QAS in \secref{approxtwotar}.

\subsection{Properties of the QAS}
\label{sec:propQAS}

\begin{figure}[!ht]
\centering
\includegraphics[width=0.5\linewidth]{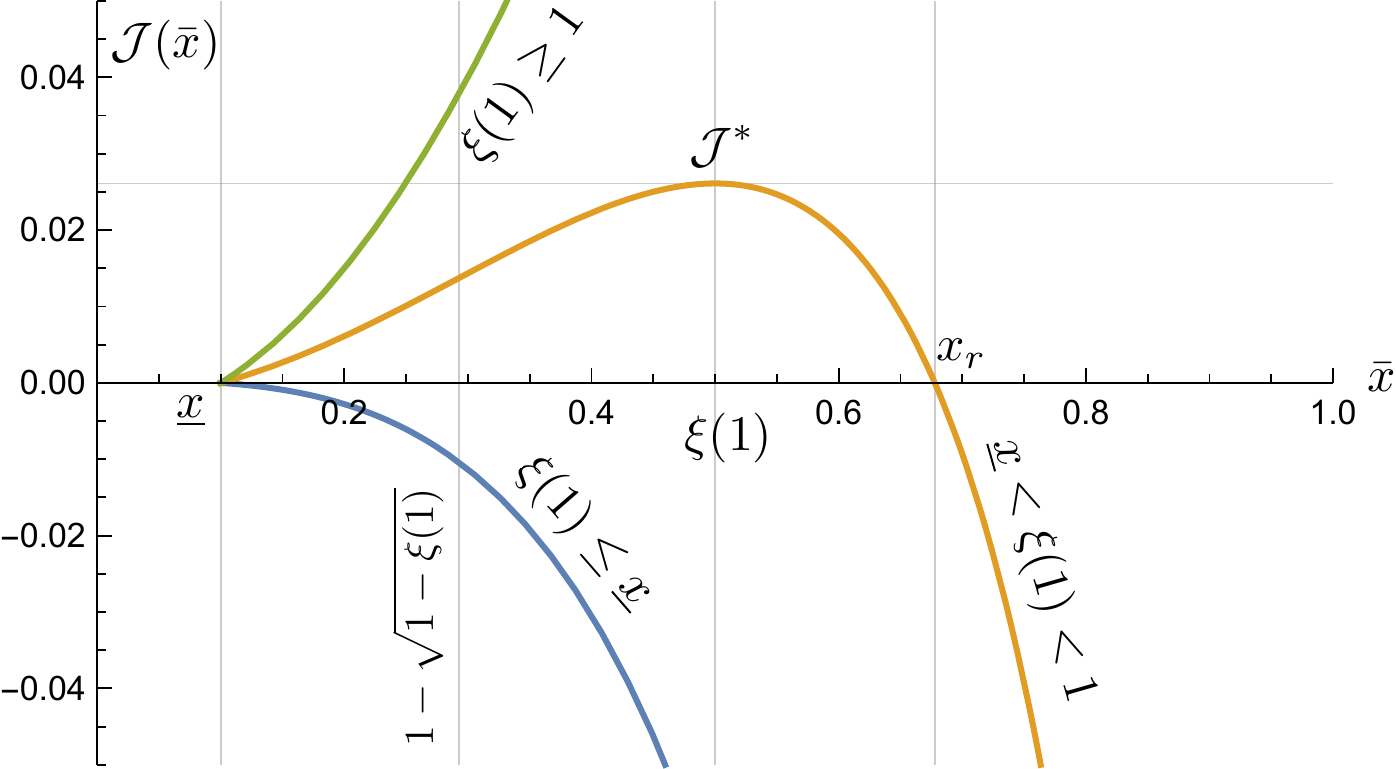}
\caption{\secref{propQAS}: Illustration of the aggregate cost born by the provider (\prpref{qasprop}) under the quickest adoption subsidy (QAS). The three curves show $\Jmc(\xu)$ vs.\ $\xu \in [\xl,1]$ for $\xi(1) \in \{0,1/2,1\}$, with $\xl = 1/10$.  The curve for $\xi(1) \in [\xl,1]$ has an inflection point at $1-\sqrt{1-\xi(1)}$, achieves its maximum $\Jmc^*$ at $\xi(1)$, and has a root at $x_r$.}
\label{fig:qascost}
\end{figure}

Quickest adoption is achievable by finite-cost subsidies under the assumption below.
\begin{assumption}
\label{ass:minaff}
There exists a {\em finite minimum service affinity} $\rhol$ with $\bar{F}_A(\rhol) = 1$.
\end{assumption}
The rationale behind this assumption is that it ensures $\rho(\chi)$ is finite at $\chi=1$, and therefore $u$ under the TTAS is finite, as the following definition makes clear.  Observe the normal affinities used in \figref{twotarmdyn} and \figref{twotarcostdur} do not satisfy this assumption.  
\begin{definition}
\label{def:qadsub}
Quickest adoption subsidies (QAS) are TTAS with $\chi = 1$.  They have finite instantaneous cost $u = c - \rhol - e x$ if $\rhol \equiv \rho(1)$ is finite, as is true under \assref{minaff}.  $\square$
\end{definition}
We view the target adoption level $\xu$ as the parameter of interest for a QAS, and establish properties of the aggregate cost $J[u]$ as a function of $\xu$ in the following proposition.
\begin{proposition}
\label{prp:qasprop}
The QAS has aggregate cost
\begin{equation}
\label{eq:QASaggcost}
\frac{\gamma}{e} J[u] = \Jmc(\xu) \equiv (1-\xi(1))((\xu - \xl) - \tau(1)) + \frac{1}{2}(\xu^2 - \xl^2) 
\end{equation}
where $\xi(1) = (c-\rhol)/e$ and $\tau(1) \equiv \log \frac{1-\xl}{1-\xu}$.  Consider $\Jmc(\xu)$ over $[\xl,1]$, noting $\Jmc(\xl) = 0$:
\begin{itemize}
\item For $\xi(1) < \xl$: $\Jmc(\xu)$ is negative and concave decreasing over $[\xl,1]$;
\item For $\xi(1) \in [\xl,1]$:  $\Jmc(\xu)$ is convex increasing over $[\xl,1-\sqrt{1-\xi(1)}]$, concave increasing over $[1-\sqrt{1-\xi(1)},\xi(1)]$, with global maximum at $\xu = \xi(1)$, and concave decreasing over $[\xi(1),1]$.  Moreover, $\Jmc(\xu)$ has a unique root in $x_r \in (\xi(1),1)$ such that $\frac{\gamma}{e} J[u]$ is positive over $\xu \in [\xl,x_r]$ and negative over $\xu \in [x_r,1]$.
\item For $\xi(1) > 1$: $\Jmc(\xu)$ is positive and convex increasing over $[\xl,1]$. $\square$
\end{itemize}
\end{proposition}
\begin{proof}
Specializing \prpref{twotargetaggcost} to $\chi=1$ yields \eqref{eq:QASaggcost}.  The two derivatives are
\begin{equation}
\dd{\xu} \Jmc(\xu) = \frac{\xu(\xi(1)-\xu)}{1-\xu}, ~ 
\ddt{\xu} \Jmc(\xu) = 1 - \frac{1-\xi(1)}{(1-\xu)^2}
\end{equation}
The three regimes follow by the signs of the two derivatives as a function of $\xu$.
\end{proof}
\begin{remark}
\label{rem:qasprop}
The condition $\xi(1) \leq \xl$ is equivalent to $\rhol + e \xl - c \geq 0$, which is easily seen to also denote the condition that the lowest possible net utility in the absence of subsidiziation, \ie $\Vl$ satisying $V \geq \Vl$ almost surely, is nonnegative, (since $\Vl \equiv \rhol - \nu(\xl,0)$).  Thus $\xi(1) < \xl$ represents a service where the {\em entire} set of potential users has a positive net utility at all adoption levels $x \geq \xl$ without subsidization.  The condition $\xi(1) \geq 1$ is equivalent to $\rhol + e 1 - c \leq 0$, which asserts that the maximum utility (over $x$) of the minimum affinity user is negative in the absence of subsidization.  Thus, the service provider desiring quickest adoption dynamics must {\em always} subsidize the service, even at near-full adoption where the externality is strongest, to ensure the lowest affinity users have positive net utility. $\square$
\end{remark}
\figref{qascost} illustrates the three behaviors from \prpref{qasprop}.  The curve for $\xi(1) \in [\xl,1]$ is positive (negative) for $\xu < x_r$ ($\xu > x_r$), meaning the provider incurs a net cost for moderate target adoption levels, but reaps a net profit for higher adoption levels.  The difference is on account of the provider being able to recoup incurred expenses in jumpstarting the adoption of the service once the externality at high adoption levels allows the use of a negative subsidy while preserving the quickest adoption property.

\subsection{Approximate QAS}
\label{sec:approxtwotar}

\begin{figure}[!ht]
\centering
\includegraphics[width=0.5\linewidth]{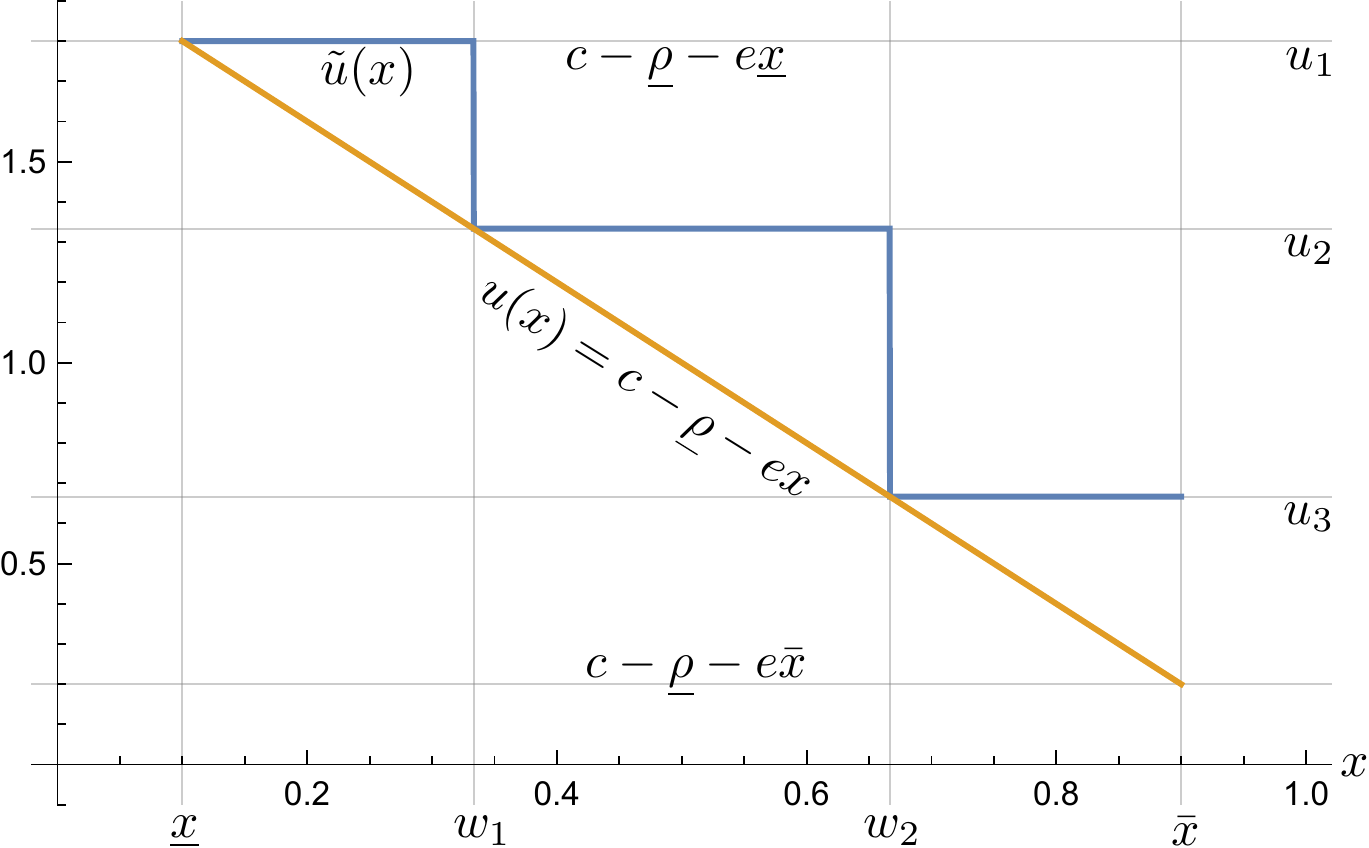}
\caption{\secref{approxtwotar}: An approximate quickest adoption subsidy (AQAS) schedule $\tilde{u}(x)$ from \defref{approxtwotar}, for $k=2$ with $w_1 = 1/3$ and $w_2 = 2/3$.  The other parameters are $\xl=1/10$, $\xu=9/10$, $c=2$, $e=2$, $\rhol=0$. }
\label{fig:approxQAS1}
\end{figure}

The TTAS, including the special case of QAS, has the drawback of requiring the provider  instantaneously adjust the subsidy amount to track the adoption level, according to $u(x(t)) = c - \rho - ex(t)$.  This may be unrealizable or undesirable for practical service deployments, which motivates the following discussion of {\em approximate} QAS, hereafter denoted AQAS, where $u$ is updated {\em discretely} at each of $k$ target intermediate adoption levels.  
\begin{definition}
\label{def:approxtwotar}
The AQAS with parameters $(\xu,w)$ employs $k$ intermediate targets $w = (w_1,\ldots,w_k)$, where $\xl = w_0 < w_1 < \cdots < w_k < w_{k+1} = \xu$.  The subsidy is set according to $\tilde{u}(x) = \tilde{u}_{i(x)}$, where $i(x) = \lceil (k+1) x \rceil \in [k+1]$ is the index at adoption level $x$, $\lceil \cdot \rceil$ is the ceiling function, and the vector $(\tilde{u}_1,\ldots,\tilde{u}_{k+1})$ has components
\begin{equation}
\tilde{u}_i \equiv c - \rhol - e w_{i-1}, ~ i \in [k+1].
\end{equation}
\end{definition}
As illustrated in \figref{approxQAS1}, the AQAS subsidy schedule $\tilde{u}(x)$ is a piecewise-constant function of $x$ that is equal to the {\em actual} QAS subsidy schedule $u(x)$ at points $w_i$, for $i \in \{0,\ldots,k\}$, and exceeds $u(x)$ at all other points.  The key property of the approximate schedule is that the excess subsidy ensures positive net utility for all potential users, and therefore quickest adoption.  That is, $\tilde{u}(x) \geq u(x)$ ensures $\nu(x,\tilde{u}(x)) \leq \nu(x,u(x)) = \rhol$, and thus $\bar{F}_A(\nu(x,\tilde{u}(x))) \geq \bar{F}_A(\rhol) = 1$.  In essence, the AQAS incurs an additional aggregate cost in exchange for a simpler subsidy schedule.

\prpref{approxQAScost} gives the aggregate cost of an AQAS.
\begin{proposition}
\label{prp:approxQAScost}
The (normalized) aggregate cost $\frac{\gamma}{e} J[u]$ of an AQAS with intermediate targets $w$ is
\begin{equation}
\Jmc(w) = \sum_{i=0}^k w_i (w_{i+1} - w_i)  - \sum_{i=0}^k w_i \log \frac{1-w_i}{1-w_{i+1}} + \xi(1)\left( \log \frac{1-\xl}{1-\xu} - (\xu-\xl) \right), \label{eq:approxQAScost}
\end{equation}
where $\Jmc(w) = \frac{\gamma}{e} J[u]$ emphasizes the dependence on $w$.  The necessary condition for $w$ to be extremal, $\nabla \Jmc(w) = 0$, is the system of $k$ equations
\begin{equation}
\label{eq:appQASgrad}
(w_{i+1}-w_i) - (w_i - w_{i-1}) + \frac{w_i - w_{i-1}}{1-w_i} - \log \frac{1-w_i}{1-w_{i+1}} = 0,
\end{equation}
for each $i \in [k]$.  The Hessian, $\nabla^2 \Jmc(w)$, is tridiagonal with components
\begin{equation}
\ddt{w_i} \Jmc(w) = \frac{2w_i}{1-w_i} + \frac{w_i - w_{i-1}}{(1-w_i)^2} > 0, ~ 
\ddtt{w_i}{w_{i+1}} \Jmc(w) = \frac{-w_{i+1}}{1-w_{i+1}} < 0
\end{equation}
and as such $\Jmc(w)$ is convex for $k=1$. $\square$
\end{proposition}
\begin{proof}
Substitute $\tilde{u}(x)$ into \eqref{eq:subfuncostalt} and rearrange as \eqref{eq:approxQAScost}: $\Jmc(w) = $
\begin{equation}
\frac{1}{e} \int_{\xl}^{\xu} \frac{x \tilde{u}(x)}{1-x} \drm x = \frac{1}{e}\sum_{i=1}^{k+1} \tilde{u}_i \int_{w_{i-1}}^{w_i} \frac{x }{1-x} \drm x = \sum_{i=1}^{k+1} (\xi(1)-w_{i-1}) \left( \log \frac{1-w_{i-1}}{1-w_i} - (w_i - w_{i-1}) \right) \nonumber 
\end{equation}
The gradient $\nabla \Jmc(w)$ and Hessian are obtained by differentiation of \eqref{eq:approxQAScost} and \eqref{eq:appQASgrad}, respectively.  The convexity for $k=1$ follows from the sign of $\ddt{w_i} \Jmc(w)$.
\end{proof}
\begin{remark}
\label{rem:ATASJconvex}
Numerical investigation suggests $\Jmc(w)$ is in fact convex in $w$ for general $k$, not just $k = 1$, but we have thus far been unable to prove this.  The well-known sufficient condition for convexity (\ie positive definiteness of $\nabla^2 \Jmc(w)$) is to establish that $\nabla^2 \Jmc(w)$ is diagonally dominant, and apply the Gershgorin circle theorem.  Unfortunately, $\nabla^2 \Jmc(w)$ is {\em not} diagonally dominant, as may be seen for the case $k=2$:
\begin{equation}
\nabla^2 \Jmc(w) = \left[ \begin{array}{cc}
\frac{2 w_1}{1-w_1} + \frac{w_1 - \xl}{(1-w_1)^2} & -\frac{w_2}{1-w_2} \\
-\frac{w_2}{1-w_2} & \frac{2 w_2}{1-w_2} + \frac{w_2 - w_1}{(1-w_2)^2} 
\end{array} \right]
\end{equation}
Diagonal dominance for the first row requires 
\begin{eqnarray}
0 < \ddt{w_1} - \left| \ddtt{w_1}{w_2} \right|, ~ 
0 < \frac{2 w_1}{1-w_1} + \frac{w_1 - \xl}{(1-w_1)^2} - \frac{w_2}{1-w_2}
\end{eqnarray}
but the latter is false for, \eg $(\xl,w_1,w_2) = (0,1/2,4/7)$, where the right side is $-2/7$. $\square$
\end{remark}
For $k=1$ the convexity of the AQAS aggregate cost in the intermediate target $w_1$ may be understood as follows (recall \figref{qascost}).  If $w_1$ is near $\xl$ and far from $\xu$ then the first subsidy level $\tilde{u}(x) = c - \rhol - e \xl$ holds for a smaller range of $x$, but the second subsidy level $\tilde{u}(x) = c - \rhol - e w_1$ holds for a larger range of $x$, and the reverse is true for $w_1$ near to $\xu$ and far from $\xl$.  The optimal $w_1^*$ makes the best tradeoff among these two costs.  

\begin{figure}[!ht]
\centering
\includegraphics[width=\linewidth]{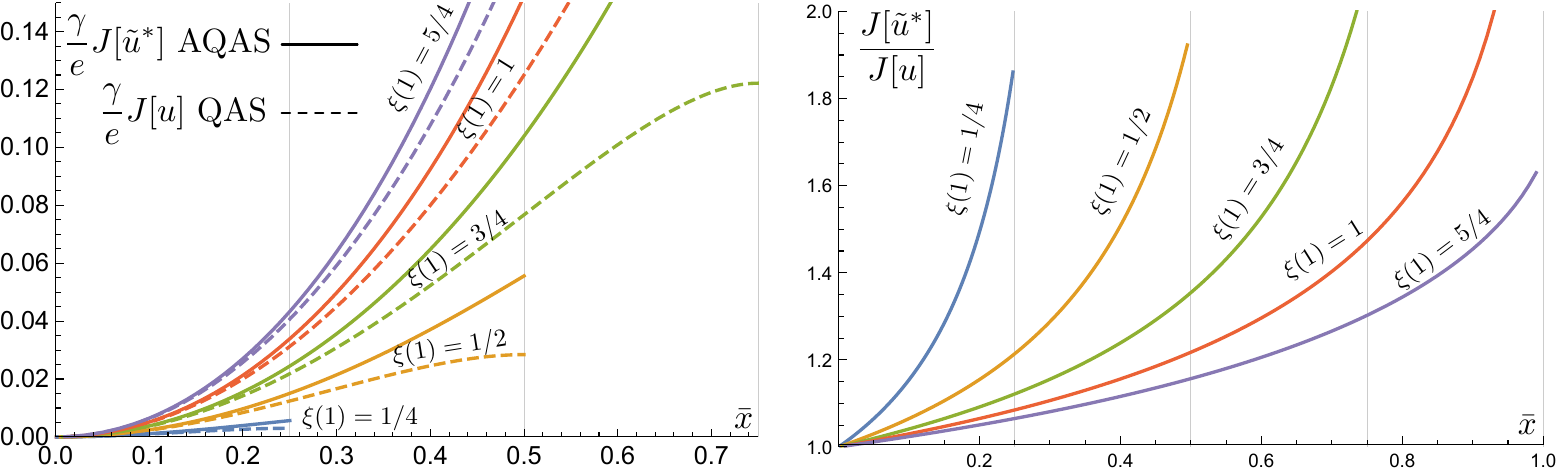}
\caption{(\secref{approxtwotar}) Left: normalized aggregate costs $\frac{\gamma}{e} J[u]$ and $\frac{\gamma}{e} J[\tilde{u}^*]$ of the QAS (\prpref{qasprop}, dashed) and the approximate QAS (AQAS, \prpref{approxQAScost}, solid) using an optimized single intermediate target ($k=1$) solving \eqref{eq:appQASgrad}, vs.\ the target adoption level $\xu$ for various values of $\xi(1) \in \{1/4,\ldots,5/4\}$.  Right: cost ratio $J[\tilde{u}^*]/J[u]$ vs.\ $\xu$, showing cost inefficiency of AQAS over QAS.}
\label{fig:qad}
\end{figure}

Selecting $w_1^*$ as a function of $(\xl,\xu)$ via \eqref{eq:appQASgrad}, yields an optimized AQAS schedule $\tilde{u}^*$, with optimized (normalized) cost $\frac{\gamma}{e} J[\tilde{u}^*]$.  Although $w_1^*$ is not expressible in closed form, and as such nor is $\tilde{u}^*$ or $J[\tilde{u}^*]$, it is nonetheless easily computed. \figref{qad} (top) plots the normalized aggregate cost of the QAS ($\frac{\gamma}{e} J[u]$) and the AQAS ($\frac{\gamma}{e} J[\tilde{u}^*]$) vs.\ the target $\xu$ for various $\xi(1)$.  The ratio $J[\tilde{u}^*]/J[u]$ (bottom) is the aggregate cost inefficiency of the optimized AQAS relative to that of QAS.  The inefficiency is increasing in $\xu$ for any $\xi(1)$.

\section{Affinities and their equilibria}
\label{sec:aff}

While optimizing the target adoption level $\xu$ for QAS and AQAS to minimize cost is a natural objective, the discussion thus far has not incorporated the {\em equilibria} associated with the {\em unsubsidized} dynamics.  In particular, if the target $\xu$ lies in the domain of attraction for a stable equilibria much lower than $\xu$, then the adoption level will immediately begin converging towards that point as soon as the subsidy is terminated, thereby ``undoing'' the forward progress achieved via the subsidy.  The natural use of the subsidy in the context of services exhibiting externalities is to use the subsidy to ``jump start'' the adoption to reach the boundary of the domain of attraction for the desired adoption level, such that, upon termination of the subsidy, the strength of the externality will maintain or increase the adoption level.  Such an objective requires knowledge of the set of equilibria, whether or not each equilibrium is stable, and how the equilibria depend upon the key model parameters: the nominal cost $c$, the externality parameter $e$, and the affinity distribution $\bar{F}_A$.  We study this question for general $\bar{F}_A$ in \secref{genaff}, the uniform distribution in \secref{uniaff}, and the normal distribution in \secref{noraff}.

\subsection{General affinities}
\label{sec:genaff}

\begin{figure}[ht]
\centering
\includegraphics[width=\linewidth]{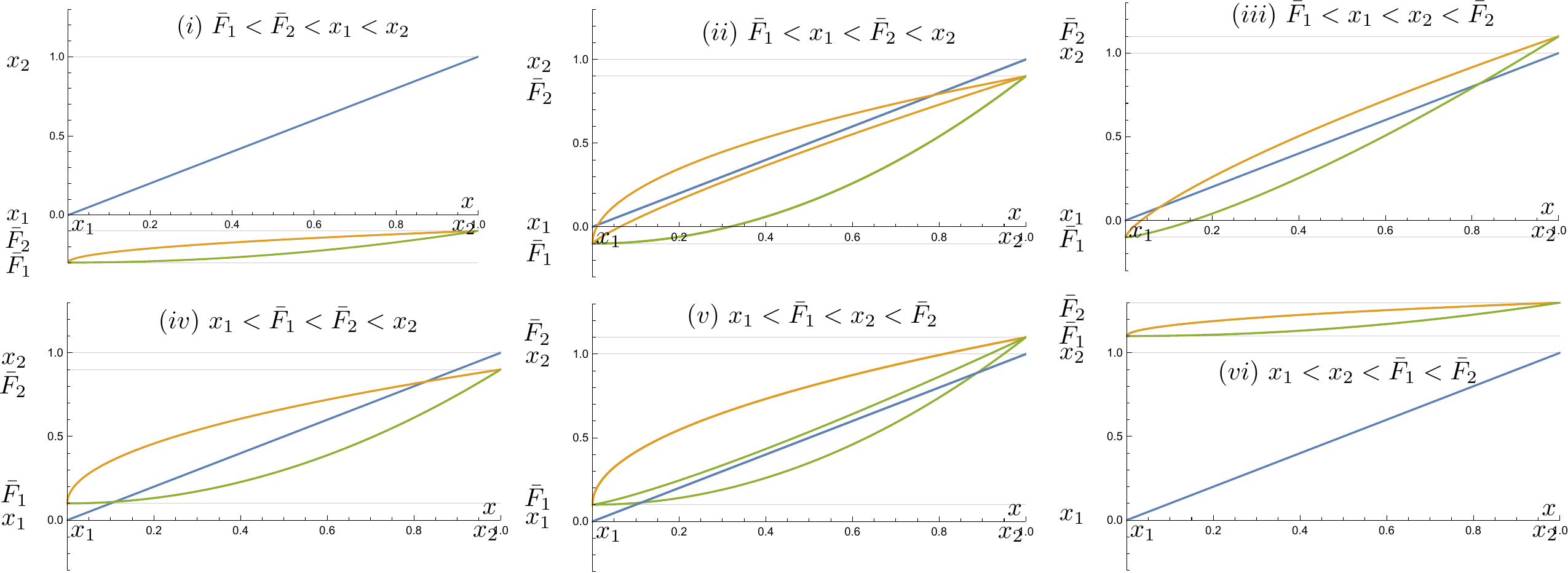}
\caption{Illustration of \prpref{numsolnconvexconcave}.  The six cases represent the six orderings of $(x_1,x_2,\bar{F}_1,\bar{F}_2)$.  Each case shows $\bar{F}_A(\nu(x,0))$, both convex (green) and concave (gold), over $x \in [x_1,x_2]$, where equilibria obey $\bar{F}_A(\nu(x,0)) = x$.}
\label{fig:numsolnconvexconcave}
\end{figure}

The support, say $\Amc$, of any continuous affinity distribution, $\bar{F}_A$, is an (possibly infinite) interval, which may be partitioned into $k$ sub-intervals, say $\Amc_1,\ldots,\Amc_k$, such that $\bar{F}_A$ is alternately convex and concave on successive intervals.  \prpref{numsolnconvexconcave}, illustrated in \figref{numsolnconvexconcave}, enumerates the number of equilibria, defined in \defref{equil}, found in sub-interval $\Amc_k$.  

Consider $x_1,x_2$, with $0 \leq x_1 < x_2 \leq 1$, as defining an interval $[x_1,x_2]$ of interest on the set of possible adoption levels $[0,1]$.  Next, define $\bar{F}_1,\bar{F}_2$, with $0 \leq \bar{F}_1 < \bar{F}_2 \leq 1$, where $\bar{F}_1 = \bar{F}_A(\nu(x_1,0))$ and $\bar{F}_2 = \bar{F}_A(\nu(x_2,0))$, and recall $\bar{F}_A(\nu(x,0))$ represents the fraction of users with positive net affinity in the absence of subsidies, \ie $u=0$.  Observe $i)$ $\bar{F}_A(a)$ is a {\em decreasing} function of $a$, but $\bar{F}_A(\nu(x,0))$ is an {\em increasing} function of $x$, and $ii)$ if $\bar{F}_A(a)$ is {\em convex} (concave) in $a$ over $[\nu(x_2,0),\nu(x_1,0)]$ then $\bar{F}_A(\nu(x,0))$ is {\em concave} (convex) in $x$ over $[x_1,x_2]$.

\begin{proposition}
\label{prp:numsolnconvexconcave}
Consider an adoption interval $[x_1,x_2]$ such that the corresponding affinity interval $[\nu(x_2,0),\nu(x_1,0)]$ is contained entirely in one of the sub-intervals $\Amc_k$ of $\bar{F}_A$, and let $[\bar{F}_1,\bar{F}_2]$ be the corresponding {\em target} adoption interval.  There are six possible orderings of the intervals $[x_1,x_2]$ and $[\bar{F}_1,\bar{F}_2]$, enumerated below, and, by construction, $\bar{F}_A(\nu(x,0))$ is either convex or concave in $x$ over $[x_1,x_2]$.  Then the number of equilibria in that interval, each solving $\bar{F}_A(\nu(x,0)) = x$, is:
\begin{equation}
\begin{array}{r|rrrr|cc}
& 1 & 2 & 3 & 4 & \cap & \cup \\ \hline
(i) & \bar{F}_1 & \bar{F}_2 & x_1 & x_2 & 0 & 0 \\
(ii) & \bar{F}_1 & x_1 & \bar{F}_2 & x_2 & 0 \mbox{ or } 2 & 0 \\
(iii) & \bar{F}_1 & x_1 & x_2 & \bar{F}_2 & 1 & 1 \\
(iv) & x_1 & \bar{F}_1 & \bar{F}_2 & x_2 & 1 & 1 \\
(v) & x_1 & \bar{F}_1 & x_2 & \bar{F}_2 & 0 & 0 \mbox{ or } 2 \\
(vi) & x_1 & x_2 & \bar{F}_1 & \bar{F}_2 & 0 & 0 
\end{array}
\end{equation}
The first column is a case label, the next four columns identify the ordering of $\{x_1,x_2,\bar{F}_1,\bar{F}_2\}$ and the last two columns indicate the number of solutions when $\bar{F}_A(\nu(x,0))$ is concave ($\cap$) or convex ($\cup$) in $x$ over $[x_1,x_2]$, respectively.  $\square$
\end{proposition}

\begin{proof}
The proof is essentially by picture.  Case $(i)$ and $(vi)$ are trivial.  Cases $(iii)$ and $(iv)$ are similar; consider case $(iii)$.  By Brouwer's fixed-point theorem there exists at least one solution.  Suppose $\bar{F}_A(\nu(x,0))$ is concave in $x$, and let $x^*$ denote the smallest element in the set of solutions.  This ensures $x^*$ is the unique solution since $\bar{F}_A(\nu(x,0))$ lies above the chord connecting $(x^*,x^*)$ and $(x_2,\bar{F}_2)$, and this chord lies above the chord connecting $(x^*,x^*)$ and $(x_2,x_2)$, and thus there can be no solutions in $[x^*,x_2]$.  Suppose $\bar{F}_A(\nu(x,0))$ is convex in $x$, and let $x^*$ denote the largest element in the set of solutions.  This ensures $x^*$ is the unique solution since $\bar{F}_A(\nu(x,0))$ lies below the chord connecting $(x_1,x_1)$ and $(x^*,x^*)$, and this chord lies below the chord connecting $(x_1,x_1)$ and $(x^*,x^*)$, and thus there is no solution in $[x_1,x^*]$.

Cases $(ii)$ and $(v)$ are similar; consider case $(ii)$.  If $\bar{F}_A(\nu(x,0))$ is convex in $x$ then there are no equilibria in $[x_1,x_2]$ since $\bar{F}_A(\nu(x,0))$ lies below the chord connecting $(x_1,\bar{F}_1)$ and $(x_2,\bar{F}_2)$, and this chord lies below the chord connecting $(x_1,x_1)$ and $(x_2,x_2)$. Suppose next $F_U$ is concave. Similar arguments establish the set of equilibria to be either $0$ or $2$ when $\bar{F}_A(\nu(x,0))$ is concave in $x$.
\end{proof}

Although the number of equilibria within each subinterval $\Amc_k$ of $\Amc$ is between zero and two, there is no limit on the number of such intervals for an arbitrary distribution $\bar{F}_A$, and as such it is difficult to develop a general theory.  We therefore address the concrete examples of uniform and normal affinities in \secref{uniaff} and \secref{noraff}, respectively.

\subsection{Uniform affinities}
\label{sec:uniaff}

In this subsection we suppose there exists $\al,\au$ with $\al \leq \au$ and that user affinities are uniformly distributed over $[\al,\au]$, \ie $A \sim \mathrm{Uni}[\al,\au]$.  Let $W \sim \mathrm{Uni}[0,1]$ be a standard uniform random variable with CCDF $\bar{F}_W(w)$.  We $i)$ characterize the set of equilibria $\Xmc$ and stable equilibria $\bar{\Xmc}$, and $ii)$ explicitly solve the unsubsidized ($u=0$) AD $x(t)$ in \eqref{eq:adoptdyn}.
  
Specializing the general AD \eqref{eq:adoptdyn} to the uniformly distributed affinities, standardizing $A$ via $W = (A-\al)/(\au-\al) \sim \mathrm{Uni}[0,1]$, and writing $c'$ for $c$ and $e'$ for $e$ yields $f(x,0) + x = $
\begin{equation}
\Pbb(A > c'-e' x) 
= \Pbb \left(\frac{A-\al}{\au-\al} > \frac{c'-e'x - \al}{\au-\al} \right)  
= \bar{F}_W \left(\frac{c'-\al}{\au-\al} - \frac{e'}{\au-\al}x \right) 
= \bar{F}_W(c - e x)
\end{equation}
where $c=(c'-\al)/(\au-\al)$ and $e = e'/(\au-\al)$.  Thus, there is no loss in generality in restriction to the case $\al=0$ and $\au=1$ since, for any $(c',e',\al,\au)$ tuple, the model $(c,e,0,1)$ is equivalent.  Because of this equivalence we henceforth assume $A \sim \mathrm{Uni}[0,1]$.  

\begin{figure}[!ht]
\centering
\includegraphics[width=\linewidth]{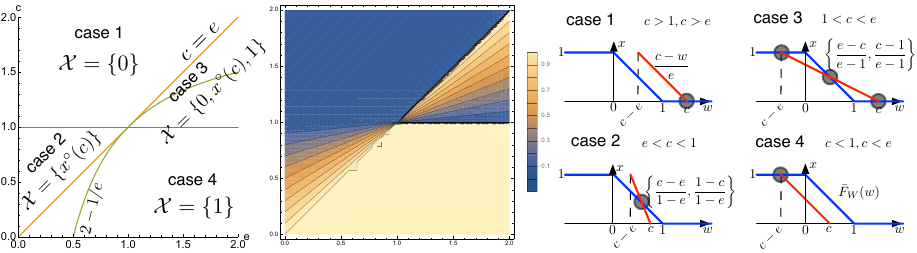}
\caption{(\secref{uniaff}) Left: the set of equilibria $\Xmc$ on the $(e,c)$ plane under uniformly distributed affinities.  Middle: $x^{\circ}$ in the region $1<c<e$.  Right: illustration of the four possible intersections of $\bar{F}_W(w)$ with the line segment $(c-w)/e$ over $w \in [c-e,e]$; black circles denote equilibria.}
\label{fig:uniaffeqmap}
\end{figure}

Our first result gives the equilibria $\Xmc$ (Def.~\ref{def:equil}) as a function of $(e,c)$, shown in Fig.~\ref{fig:uniaffeqmap}.  The (unstable) equilibria $x^{\circ}$ below is, when $x^{\circ} \in (0,1)$, the boundary between the domains of attraction to the stable equilibria at $0$ and $1$:
\begin{equation}
\label{eq:xcircc}
x^{\circ} \equiv \frac{c-1}{e-1}.
\end{equation}
\begin{proposition}
\label{prp:uniaffnumeq}
The equilibria under uniform affinities are:
\begin{equation}
\begin{array}{lll}
\mbox{case} & \mbox{region} & \Xmc \\ \hline
1 & c > 1, c > e & \{0\} \\
2 & e < c < 1 & \{x^{\circ}\} \\
3 & 1 < c < e & \{0,x^{\circ},1\} \\
4 & c < 1, c < e & \{1\}
\end{array}
\end{equation}
Besides these main cases, there are ``edge cases'': 
$i)$ $\Xmc = [0,1]$ for $c = e = 1$,
$ii)$ $\Xmc = \{0\}$ for $e<c=1$,
$iii)$ $\Xmc = \{1\}$ for $c=e <1$,
$iv)$ $\Xmc = \{0,1\}$ for $1<c=e$ and $c=1<e$.
All equilibria are stable, \ie $\Xmc = \bar{\Xmc}$, except $x^{\circ}$ in case 3. $\square$
\end{proposition}
\begin{proof}
See the far left of Fig.~\ref{fig:uniaffeqmap}. From Def.~\ref{def:equil}, $\Xmc$ are the $x$-coordinates of the intersections of $\bar{F}_W(w)$ and the line segment $(c-w)/e$ on the $(w,x)$ plane.  As the adoption level $x$ obeys $0 \leq x \leq 1$, it follows that $c-e \leq w \leq c$.  It is clear that $\Xmc \neq \emptyset$ for all $(e,c)$. There are four cases for this intersection, as shown in the figure.
\end{proof}

Define 
\begin{equation}
\label{eq:nosubuniafftx1}
\hat{t}(x|\tl,\xl) \equiv \tl + \frac{1}{(e-1) \gamma} \log \left(\frac{x-x^{\circ}}{\xl-x^{\circ}} \right)
\end{equation}
as the time $t$ at which $x(t)$ solving $\dot{x}(t) = \gamma f(x,0)$ reaches $x$ (with $x(\tl)=\xl$), for $(c-1)/e \leq x \leq c/e$, and  
\begin{equation}
\label{eq:nosubuniaffT10}
\hat{T}_1(\xl) \equiv \hat{t}\left(\left.\frac{c-1}{e}\right|0,\xl\right), ~  
\hat{T}_0(\xl) \equiv \hat{t}\left(\left.\frac{c}{e}\right|0,\xl\right) 
\end{equation}
as the time durations required to reach $(c-1)/e$ and $c/e$, respectively, starting from $\xl$, assuming such times are finite.

\begin{proposition}
\label{prp:nosubAD}
The AD under uniform affinities when $1 < c < e$ (case $3$) are: 
\begin{equation}
\label{eq:adonecolcase3}
x_3(t|\tl,\xl) = 
\left\{ 
\begin{array}{llllllll}
\xl \erm^{-\gamma(t-\tl)} & & & \xl & \leq & \frac{c-1}{e} & & \\
x^{\circ} + (\xl - x^{\circ}) \erm^{(e-1) \gamma (t-\tl)} & \frac{c-1}{e} & \leq & \xl & \leq & x^{\circ} & \mbox{ and } & t - \tl \leq \hat{T}_1(\xl)\\
\frac{c-1}{e} \erm^{-\gamma(t-\tl-\hat{T}_1(\xl))} & \frac{c-1}{e} & \leq & \xl & \leq & x^{\circ} & \mbox{ and } & t - \tl > \hat{T}_1(\xl) \\
x^{\circ} + (\xl - x^{\circ}) \erm^{(e-1)  \gamma (t-\tl)} & x^{\circ} & \leq & \xl & \leq & \frac{c}{e} & \mbox{ and } & t - \tl \leq \hat{T}_0(\xl)\\
1 - \left(1-\frac{c}{e} \right) \erm^{-\gamma (t-\tl-\hat{T}_0(\xl))} & x^{\circ} & \leq & \xl & \leq & \frac{c}{e} & \mbox{ and } & t - \tl > \hat{T}_0(\xl) \\
1 - (1-\xl) \erm^{-\gamma(t-\tl)} & \frac{c}{e} & \leq & \xl & & & &
\end{array} \right.
\end{equation} 
The first (last) three subcases hold for $\xl \lessgtr x^{\circ}$, so $x_3(t) \to 0 ~ (1)$ as $t \to \infty$, respectively.  $\square$
\end{proposition}

\begin{figure}[!ht]
\centering
\includegraphics[width=\linewidth]{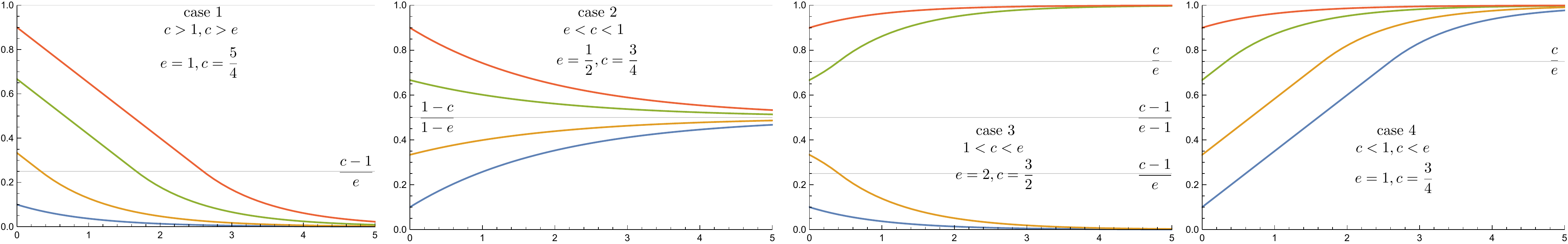}
\caption{(\secref{uniaff}): the AD $x(t|\tl,\xl)$ vs.\ $t$ (with $\tl = 0$ and $\xl \in \{1/10, 1/3, 2/3, 9/10\}$) under uniform affinities, for the four cases in \prpref{nosubAD} and \prpref{nosubAD2}.  The only case with multiple equilibria is case $3$.}
\label{fig:nosubuniaff}
\end{figure}

The AD for the remaining three cases from \prpref{uniaffnumeq} are in \prpref{nosubAD2} in \appref{proofs}, which also contains the proof of \prpref{nosubAD}.  \figref{nosubuniaff} illustrates the AD for all four cases.  

\begin{remark}
\label{rem:whycase3}
As cases 1, 2, and 4 have only one equilibrium, there is no possibility for a finite-duration subsidy to change the equilibrium adoption level.  By contrast, such a change {\em is} possible under case 3 (where $\bar{\Xmc} = \{0,1\}$) provided the initial adoption level $\xl$ lies below the boundary $x^{\circ}$ between the two domains of attraction of the two stable equilibria, $0$ and $1$, \ie $\xl < ~ (>) ~ x^{\circ}$ ensures $x(t) \to 0 ~ (1)$, respectively, as $t \to \infty$.  This boundary is the natural target adoption level, \ie $\xu = x^{\circ}$, for a subsidy, since the strength of the externality at adoption level above $x^{\circ}$ will henceforth drive the adoption level towards one without requiring a subsidy. $\square$ 
\end{remark}

\subsection{Normal affinities}
\label{sec:noraff}

Now let affinities be normally distributed, $A \sim \Nmc(\mu,\sigma)$, and first introduce some notation.  Let $Q(z) \equiv \Pbb(Z > z)$ be the CCDF of $Z \sim \Nmc(0,1)$ a standard normal, $Q^{-1}(q)$ its inverse, $q(z) \equiv \frac{\drm}{\drm z} Q(z)$ the standard normal PDF, and 
\begin{equation}
\label{eq:normalpdfinv}
q^{-1}(p) \equiv \sqrt{-2 \log(\sqrt{2\pi} p)}
\end{equation}
the (positive) inverse of $q(z)$ for $p \in (0,1/\sqrt{2\pi}]$, \ie $z=q^{-1}(p) > 0$ obeys $q(z) = p$.  

\begin{figure}[!ht]
\centering
\includegraphics[width=\linewidth]{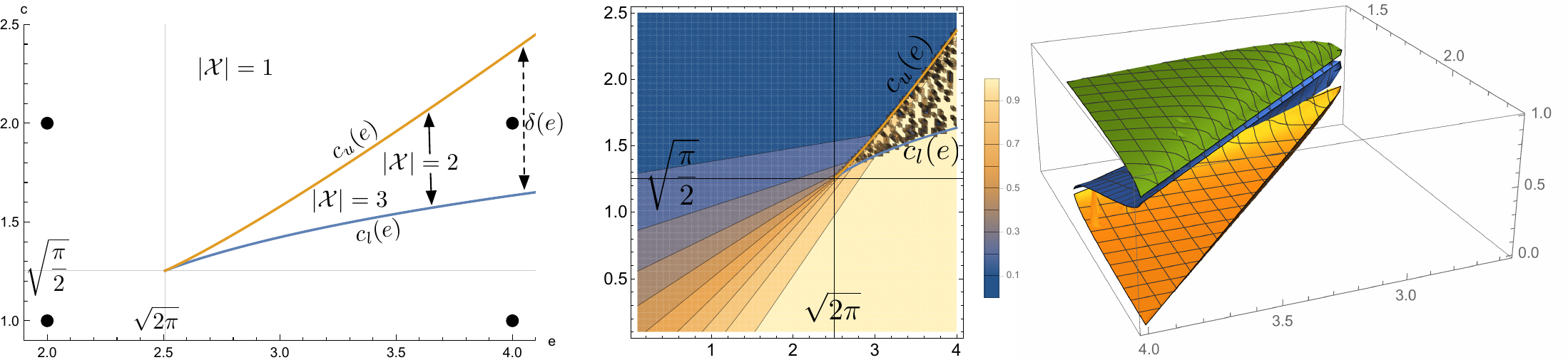}
\caption{(\secref{noraff}): numerically computed equilibria on the $(e,c)$ plane. Left: the number of equilibria $|\Xmc|$ on the $(e,c)$ plane under normally distributed affinities.  Multiple equilibria are only possible for sufficiently large externalities $e > \sqrt{2\pi}$ and intermediate costs, $c \in [c_l(e),c_u(e)]$.  The four circles are the four $(e,c)$ pairs for the AD in \figref{normaffAD}.
Middle: contour plot of the equilibrium adoption level $\bar{x}_1$; the visual occlusions in the region $e > \sqrt{2\pi}$ and $c \in [c_l(e),c_u(e)]$ are due to numerical instability from the multiple equilibria in that region.  Right: 3D-plot of the three equilibria $\bar{x}_1 < \bar{x}_2 < \bar{x}_3$ in that region.}
\label{fig:normaffnumeq}
\end{figure}

We now show that it suffices to consider $A \sim \Nmc(0,1)$.  Suppose $c',e'$ are the nominal cost and externality.  Then, via \eqref{eq:adoptdyn}, standardizing $A$ into $Z$ via $Z = (A-\mu)/\sigma$, and defining $c=(c'-\mu)/\sigma$ and $e = e'/\sigma$ we obtain: 
\begin{equation}
f(x,0) +x 
= \Pbb(A > c' - e' x)
= \Pbb \left(\frac{A-\mu}{\sigma} > \frac{c'- e'x - \mu}{\sigma}\right) 
= Q \left( \frac{c'-\mu}{\sigma} - \frac{e'}{\sigma} x\right)
= Q(c - e x).
\end{equation}
Thus there is no loss in generality in restricting attention to the case $\mu = 0$ and $\sigma = 1$ since, for any $(c',e',\mu,\sigma)$ tuple, we can obtain an equivalent model $(c,e,0,1)$.  Because of this equivalence we henceforth assume $A \sim \Nmc(0,1)$.  Although we will give our results in terms of the adoption level $x$, it is often simpler to work with the linear reparameterization $z = c - ex$, with $\dot{z} = - e \dot{x}$.  Our first result is on the number of equilibria as a function of $(e,c)$, shown in \figref{normaffnumeq}.  Define, for $e > \sqrt{2\pi}$,
\begin{equation}
c_l(e) \equiv e Q(q^{-1}(1/e))+q^{-1}(1/e), ~ 
c_u(e) \equiv e (1-Q(q^{-1}(1/e)))-q^{-1}(1/e)
\label{eq:normaffeqcritc}
\end{equation}
Note $c_l(e) \leq c_u(e)$, and the interval $[c_l(e),c_u(e)]$ has width
\begin{equation}
\label{eq:normaffdelta}
\delta(e) = e(1-2Q(q^{-1}(1/e)))-2q^{-1}(1/e).
\end{equation}
\begin{proposition}
\label{prp:normaffnumeq}
The equilibria set $\Xmc$ for normally distributed affinities has cardinality: $i)$ $|\Xmc|=3$ for $e > \sqrt{2\pi}$ and $c \in (c_l(e),c_u(e))$, $ii)$ $|\Xmc|=2$ for $e > \sqrt{2\pi}$ and $c \in \{c_l(e),c_u(e)\}$, and $iii)$ $|\Xmc|=1$ otherwise. $\square$
\end{proposition}
The proof is in \appref{proofs}.  The functions $c_l(e),c_u(e),\delta(e)$ in \eqref{eq:normaffeqcritc} and \eqref{eq:normaffdelta} are unbounded as $e \to \infty$, with $c_l(e)$ growing approximately like $\sqrt{2\log(e)}$, and $c_u(e),\delta(e)$ growing approximately linearly in $e$ for $e$ large.  As such, for any $e > 0$ and sufficiently large $c$ we recover $|\Xmc|=1$.  Likewise, for any $c > 0$ and sufficiently large $e$ we recover $|\Xmc|=1$.

\begin{figure}[!ht]
\centering
\includegraphics[width=\linewidth]{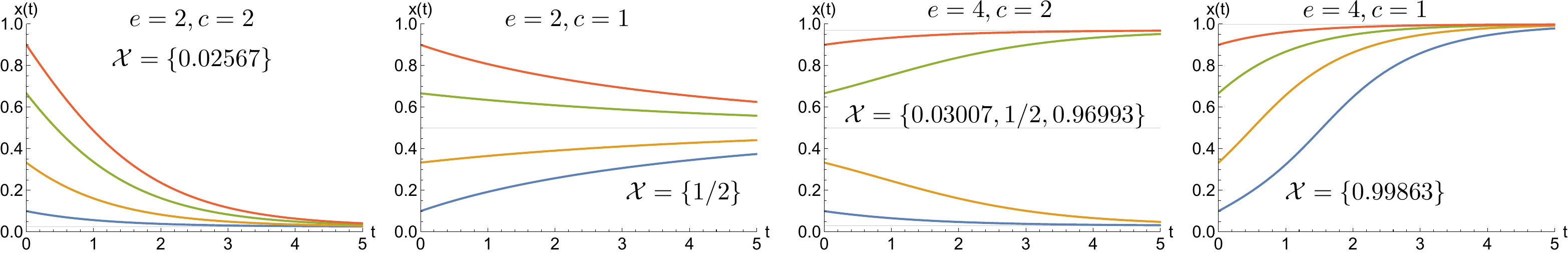}
\caption{(\secref{noraff}): AD under normally distributed affinities for the four different $(e,c)$ pairs shown as black dots in \figref{normaffnumeq} (left), with initial conditions $\xl \in \{1/10,1/3,2/3,9/10\}$. }
\label{fig:normaffAD}
\end{figure}

Although it is difficult to explicitly and analytically characterize the exact values of the equilibria $\Xmc$ for an arbitrary point on the $(e,c)$ plane, the observations below follow from \figref{normaffeqmapproof}, \figref{normaffnumeq} and the proof of \prpref{normaffnumeq}.  Below, the equilibria are numbered as $\bar{x}_1$, $\bar{x}_1 < \bar{x}_2$, and $\bar{x}_1 < \bar{x}_2 < \bar{x}_3$, for the three cases $|\Xmc| \in \{1,2,3\}$, respectively.
\begin{enumerate}
\item There always exists at least one stable equilibrium.
\item At $(e,c) = (\sqrt{2\pi},\sqrt{\pi/2})$, $\bar{x}_1 = 1/2$.
\item For any $(0,c)$ $\bar{x}_1 = Q(c) = \Pbb(A > c)$, consistent with \secref{noext}.
\item For any $(e,0)$, $\bar{x}_1 = 1$ since $\dot{z}(t) < 0$, ensuring $z(t) \to c-e$.
\item For any $(e,c_l(e))$, $\bar{x}_1$ is stable and $\bar{x}_2 = Q(q^{-1}(1/e))$ is unstable, and for any $(e,c_u(e))$, $\bar{x}_1 = 1-Q(q^{-1}(1/e))$ is unstable and $\bar{x}_2$ is stable.
\item For any $(e,c)$ with $c_l(e) < c <  c_u(e)$, equilibria $\bar{x}_1,\bar{x}_3$ are stable and $\bar{x}_2$ is unstable, with $\bar{x}_1 < Q(q^{-1}(1/e)) < \bar{x}_2 < 1-Q(q^{-1}(1/e)) < \bar{x}_3$.
\item For any $e$, $\lim_{c \to \infty} \bar{x}_1 = 0$, and $\lim_{c \to 0} \bar{x}_1$ is the unique solution to $Q(z) = -z/e$, which is increasing in $e$.
\item For any $c$, $\lim_{e \to \infty} \bar{x}_1 = 1$ and $\lim_{e \to 0} \bar{x}_1 = Q(c)$. 
\end{enumerate}
These observations are born out in \figref{normaffnumeq}, which shows the numerically computed equilibria on the $(e,c)$ plane.  Numerically computed AD are shown in \figref{normaffAD}.

\section{A case study}
\label{sec:case}

The discrete-time finite-population AD below are motivated by the model discussion in \appref{proofs} (see proof for notation).  If the fraction $x_t$ of the finite population of size $N$ adopting the service in slot $t$ is governed by the dynamics described in \appref{proofs}, then
\begin{equation}
\label{eq:b9f34gf34}
x_{t+1} = \frac{1}{N} \sum_{i \in [N]} \sigma_{i,t+1} \mathbf{1}(V_i(x_t,u_{t+1}) > 0) + (1-\sigma_{i,t+1}) b_{i,t}.
\end{equation}
The above AD are an accurate approximation of the continuous-time infinite-population AD in \eqref{eq:9fd9034f34g} for large $N$ and small time slot duration $h$, as discussed in \appref{proofs}.  

The following is a {\em fictional} example / case study intended to illustrate a potential application of the preceding content in a (hopefully) realistic scenario.  Inspired by recent studies of American communication habits finding that ``texting is the dominant way of communicating for Americans under 50'' \cite{New2014,How2015}, a new startup named {\em Texteraction} has developed a smart phone service enabling subscribers to use text communication as a replacement for in-person interactions.  When the {\em Texteraction} app is active, profile pictures of other users located within speaking range of one another will appear on the app's ``nearby'' list for each user.  Selecting a nearby subscriber will send a ``texteraction request'' to that person's phone which, if accepted, will then initiate a text session between the two parties.  The startup selects a small town of $N=1,000$ residents as its initial market, and restricts membership to those citizens.  The startup plans for three phases: $i)$ a trial period (during which citizens are not charged for the service) for parameter estimation, $ii)$ a subsidy period to drive the subscriber base to critical mass, and $iii)$ full unsubsidized service deployment.  

{\em Parameter estimation.} The objective of the trial period is to enable the startup to estimate the service externality parameter, $e$, and the town's affinity distribution $F_A$.  All $N$ citizens participate in the trial, which is conducted over a period of $T_1=10$ weeks, with one week per stage.  In week $t \in \{1,\ldots,T_1\}$ the ``nearby'' list for citizen $i$ is restricted to only show nearby citizens listed in a randomly selected subset $\Imc_{i,t} \subseteq [N]$ of size $|\Imc_{i,t}|=t N/T_1$, i.e., a random subset of size $100$ in week one, with the size increasing by $100$ in each week, until in the last week each citizen has access to all $N$ citizens.  At the beginning of the first week, and then at the end of each week of the trial, the users are asked to rate the app, with each rating intended to reflect their perceived value or utility of the service.  The ratings are collected in the $N \times (T_1+1)$ matrix $z = (z_{i,t})$, with entry $z_{i,t}$ the app rating by citizen $i$ in week $t$ of the trial ($z_{i,0}$ the initial rating).   Suppose user utility is linear and obeys Metcalfe's law, with some noise incurred in reporting the utility.  That is, we assume the ratings take the form $z_{i,t} \equiv A_i + e x_t + \epsilon_{i,t}$, where $A_i$ the (unknown) natural affinity for the service by citizen $i$, $x_t = t/T_1$ (with $x_0 = 0$) the (known) fraction of the subscriber base in week $t$, $e$ the (unknown) externality effect, and $\epsilon_{i,t} \sim \Nmc(0,\sigma_{\epsilon})$ a sequence of iid (in both $i$ and $t$) random variables (unknown) capturing the error or noise between the ``true'' utility and the user's rating.  

The externality effect in week $t \in [T_1]$ for user $i \in [N]$, is estimated via a standard slope estimator, i.e., $\hat{e}_{i,t} \equiv (z_{i,t}-z_{i,t-1})/(x_t - x_{t-1})$, yielding $\hat{e}_{i,t} = e + T_1 (\epsilon_{i,t}-\epsilon_{i,t-1})$ under the model assumption.  The estimate of $e$ by user $i$ is defined as $\hat{e}_i \equiv \frac{1}{T_1}\sum_{t \in [T_1]} \hat{e}_{i,t}$, and the telescoping error terms yield $\hat{e}_i = e + (\epsilon_{i,T_1}-\epsilon_{i,0})$.  The user average is defined as $\hat{e} \equiv \frac{1}{N}\sum_{i \in [N]} \hat{e}_i$, and may be computed to equal $\hat{e} = e + \frac{1}{N}\sum_{i \in [N]} (\epsilon_{i,T_1}-\epsilon_{i,0})$. It follows that $\hat{e} \sim \Nmc(e,\sqrt{2/N} \sigma_{\epsilon})$, and thus $\hat{e}$ is an unbiased and consistent estimator of $e$.

The user affinity distribution $F_A$ is also measured from $z$, with $\hat{A}_{i,t} \equiv z_{i,t} - \hat{e} x_t$ the estimated affinity of user $i$ in period $t$.  Algebra yields $\hat{A}_{i,t} = A_i + (e-\hat{e})x_t + \epsilon_{i,t}$.  The estimate for user $i$ is defined as $\hat{A}_i \equiv \frac{1}{T_1}\sum_{t \in [T_1]} \hat{A}_{i,t}$, and algebra yields $\hat{A}_i = A_i +(1+1/T_1)(e-\hat{e})/2 + \frac{1}{T_1} \sum_{t \in [T_1]} \epsilon_{i,t}$.  As $\hat{e} \sim \Nmc(e,\sqrt{2/N} \sigma_{\epsilon})$ and $\epsilon_{i,t} \sim \Nmc(0,\sigma_{\epsilon})$, it follows that $\hat{A}_i \sim \Nmc (A_i,\sigma_{\epsilon} g(N,T_1))$, with $g(N,T_1) = \sqrt{(1+1/T_1)^2/(2N) + 1/T_1}$.  One might therefore expect the estimator to perform adequately in estimating $A_i$ provided the coefficient of variation is sufficiently small, \ie the estimator may perform poorly for users with small natural affinity $A_i$.  The empirical distribution $\hat{F}_A$ from  $(\hat{A}_i, i \in [N])$ estimates the affinity distribution $F_A$.

The weekly cloud hosting and cellular provider data sharing costs associated with dynamically maintaining each user's ``nearby'' list are found to scale linearly with the number of users at a cost of \$1.50 per user.  The management elects to employ a paid subscription model, passing these costs to the users, i.e., $c=1.5$.  The empirical affinity estimate distribution $\hat{F}_A$ and the estimate $\hat{e}$ are used to estimate the equilibrium $\hat{x}^{\circ}$ via $1-\hat{F}_A(c-\hat{e} \hat{x}^{\circ}) = \hat{x}^{\circ}$ (c.f.\ \defref{equil}), equivalently, $\hat{F}_A(\hat{x}^{\circ}) = (1-c/\hat{e})+\hat{x}^{\circ}/\hat{e}$.   

\begin{figure}[ht]
\centering
\includegraphics[width=\linewidth]{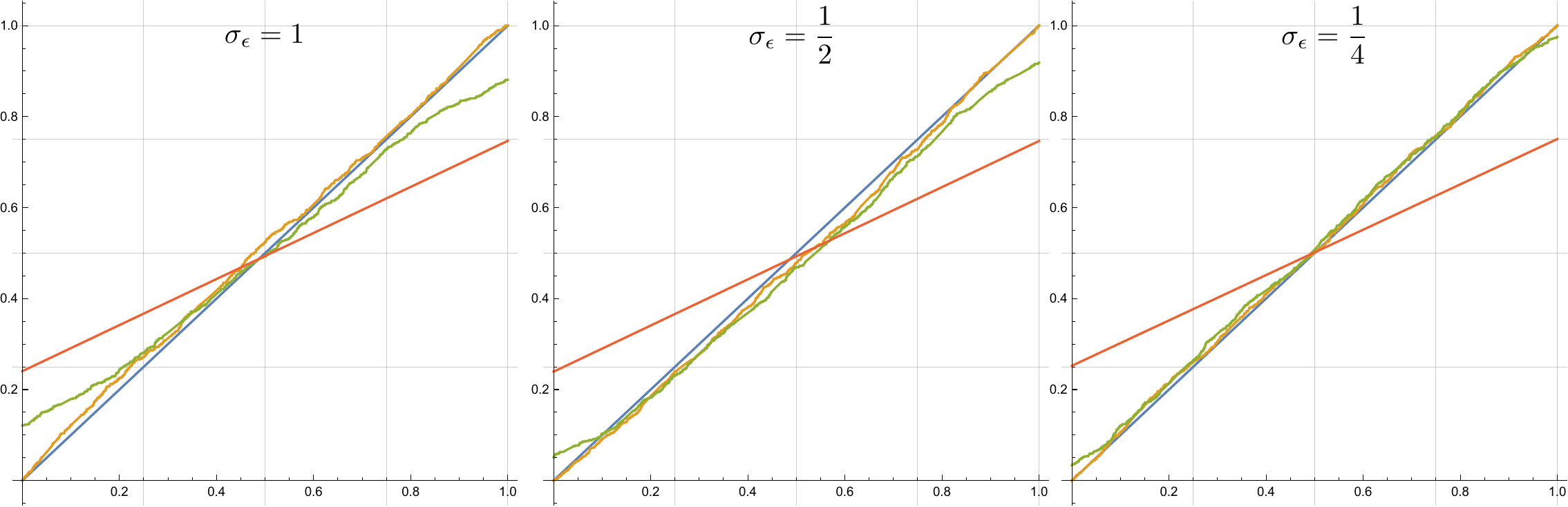}
\caption{Case study (\secref{case}): $i)$ estimated user affinities empirical distribution $\hat{F}_A(a)$ (green), $ii)$ true user affinities empirical distribution (gold), $iii)$ uniform cumulative distribution $F_A(a)$ (blue), $iv)$ line $(1-c/\hat{e})+a/\hat{e}$ (red) for estimate $\hat{x}^{\circ}$.  Measurement noise standard deviation $\sigma_{\epsilon} = 1$ (left), $1/2$ (middle), $1/4$ (right).}
\label{fig:casestudy1}
\end{figure}

Simulations of the trial period, its measurements and estimates, were performed.  Set $e=2$ and let $A_i \sim \mathrm{Uni}[0,1]$ be iid in $i \in [N]$.  Note $x^{\circ} = 1/2$ from \eqref{eq:xcircc}.  We considered three different measurement noise levels $\sigma_{\epsilon} = \{1,1/2,1/4\}$, yielding estimates $\hat{e} = \{1.975,1.972,2.006\}$ and $\hat{x}^{\circ} = \{0.500,0.497,0.500\}$, respectively, and empirical user affinity approximation distributions $\hat{F}_A$ shown in \figref{casestudy1}.  The accuracy of $\hat{F}_A$ improves as $\sigma_{\epsilon}$ is reduced, with errors in estimating small or large affinities more pronounced than for estimating intermediate affinities, as expected.  We omit due to space constraints a discussion of the estimate of $\gamma$.  As accurate measurements of $e$ and $F_A$ are seen to be possible, we henceforth assume $\gamma,e,F_A$ are now known.

{\em Subsidized deployment.} The management uses knowledge of $c,e,x^{\circ},F_A,\gamma$ to evaluate the hypothetical performance of various subsidies. The time unit is set at a week, the time slot duration is set at a day ($h=1/7$), and the subsidy phase is set to last for at most $T=90$ days.  The value of $\gamma$ is set at $1/4$, corresponding to users waiting $28$ days, on average, between subscription reassessments (c.f. \appref{proofs}).  The goal of a subsidy is to reach the fractional adoption level $\hat{x}^{\circ} = 1/2$, corresponding to $n^{\circ} = N x^{\circ} = 500$ adopters, at which point a subsidy will terminate, and the app will be self-sustaining at the nominal operating cost of $c$ per user per week.  A randomly selected initial set of $n_0 =  100$ ($x_0 = n_0/N = 0.1$) adopters are identified before the start of the subsidy period.  The actual duration of a subsidy $u$ is defined as $\bar{t}[u] \equiv \max\{t \in [T] : x_t < x^{\circ}\}$.

\begin{figure}[ht]
\centering
\includegraphics[width=0.93\linewidth]{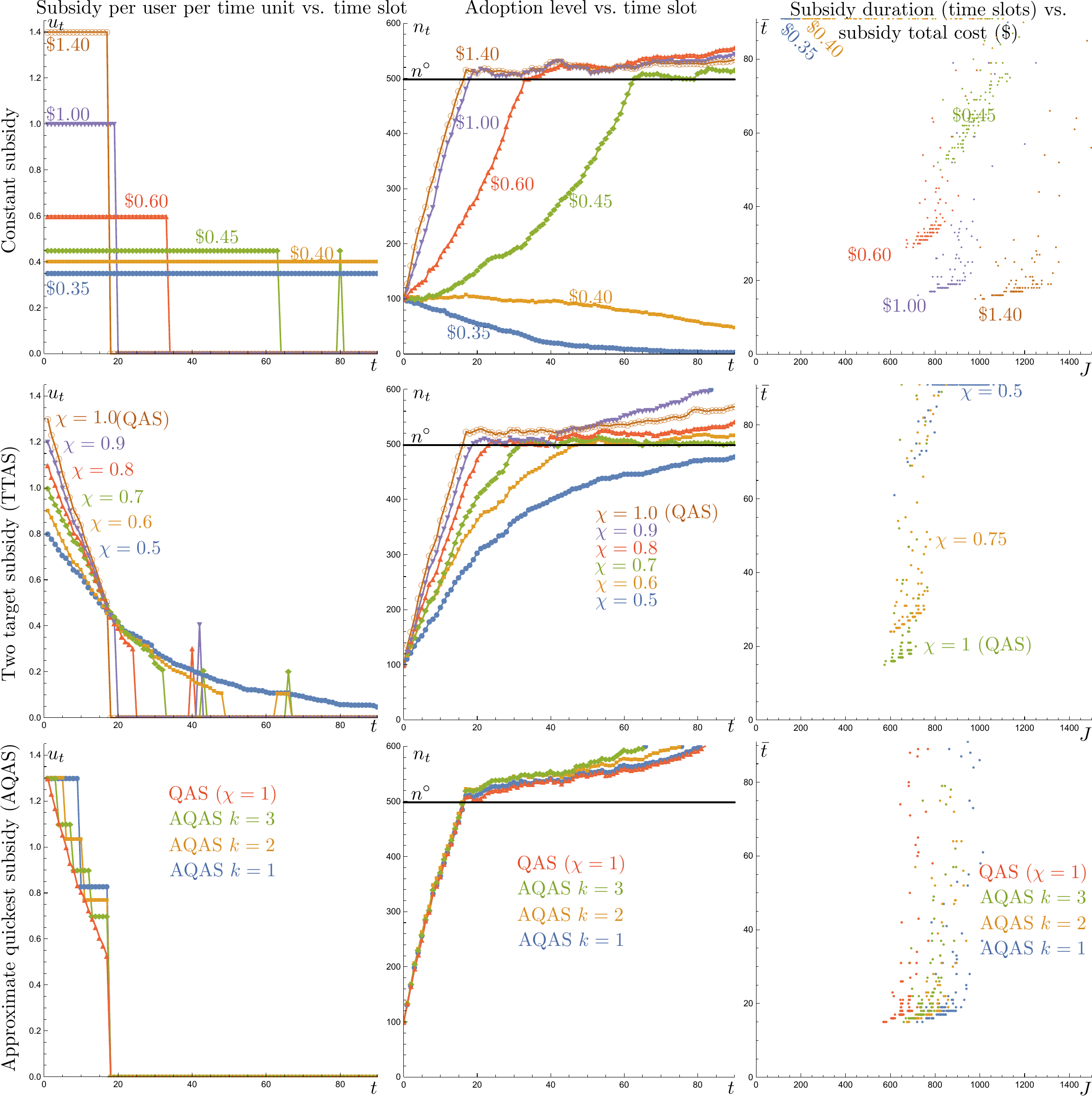}
\caption{Case study (\secref{case}) -- performance of four subsidy types.  Rows: constant subsidies (top row), TTAS and QAS (middle row), AQAS and QAS (bottom row).  Columns: sample realization of the daily per-user subsidy to the weekly service cost (left column), sample realization of the daily adoption level (middle column), scatter plot of subsidy duration vs.\ total subsidy cost (right column) for $M=100$ independent simulations.}
\label{fig:casestudy2}
\end{figure}

Four different types of subsidies are considered, with simulation results summarized in \figref{casestudy2}.  First, six different constant subsidies with subsidy $v$ while $x_t < x^{\circ}$, i.e., $u_{t+1} = v\mathbf{1}(x_t < x^{\circ})$, are evaluated, with $v \in \{\$0.35,\$0.40,\$0.45,\$0.60,\$1.00,\$1.40\}$.  Note $v$ is the subsidy of the weekly cost $c$ but is adjusted daily, and subscribed users in day $t$ are charged a daily cost of $h(c-u_t)$.  As evident from the plots, $v \in \{\$0.35,\$0.40\}$ is insufficient to change the adoption level, $v = \$0.45$ is on the boundary and depending upon the realization may or may not achieve $x^{\circ}$ by $T=90$, while $v \in \{\$0.60,\$1.00,\$1.40\}$ are sufficient to achieve the target.  There is clearly little incentive to choose $v=\$1.40$ over $v=\$1.00$.  The $(J,\bar{t})$ scatter plot shows a tradeoff in $(J,\bar{t})$ between $v \in \{\$0.60,\$1.00\}$.

Second, six TTAS subsidies were studied, with $\chi \in \{0.5,0.6,0.7,0.8,0.9,1.0\}$, and $\chi = 1.0$ the QAS.  The effect of $\chi$ on $u_t$ is shown in the left plot, with higher $\chi$ yielding a shorter duration $\bar{t}$ to reach $n^{\circ}$ (middle plot), and the $(J,\bar{t})$ scatter plot showing $\chi = 1$ (QAS) superior to both $\chi \in \{0.50,0.75\}$ in both $J$ (slight) and $\bar{t}$ (significant).

Third, three different AQAS subsidies were studied, with $k \in \{1,2,3\}$, and the QAS ($\chi =1$) included for contrast.  The intermediate adoption levels were chosen to be  $w_1^* \approx 0.336$ for $k=1$, $w = \{7/30,11/30\}$ for $k=2$, and $w=\{2/10,3/10,4/10\}$ for $k=3$ (the latter two schedules chosen to divide $[\underline{x},x^{\circ}]$ into $k+1$ equal intervals.  The AQAS achieve adoption levels and $(\bar{t},J)$ tradeoffs comparable to that of the QAS, with the scatter plot showing smaller values of $k$ incurring a slight increase in $J$, on average.  

The management elects the AQAS with $k=2$, budgeting $\$600-800$ for the cost of the subsidy, and anticipates reaching $n^{\circ} = 500$ customers within around three weeks.  

\section{Limitations}
\label{sec:limitation}

The previous section has outlined a possible methodology by which the fundamental model parameters, i.e., the externality, cost, and population affinity distribution, could conceivably be estimated through a suitable measurement campaign.  Although the measurement campaign we have described is feasible, it is not without both economic and econometric difficulties.  First, the economic costs, including monetary, labor, and service deployment delay costs, may be substantial, and may well be deemed to exceed the value of the information acquired by the measurements.  Second, the econometric difficulties are likewise significant, as reliable inference requires synchronization, or at least coordination, of the various cohorts in order to meaningfully gauge the strength of the externality at the various stages of the experiments.  The costs incurred in overcoming the substantial obsctacles required to achieve this synchronization may again be deemed too high by the service provider. 

It is worth emphasizing there are two distinct sources of ``error'' in seeking to apply diffusion adoption models to predict a ``real-world'' response: parameter error and model error.  The measurement campaign described previously seeks to accurately estimate the model parameters, but there is no assurance that the target users will, in all cases, behave in a manner even approximately consistent with the predictions of the diffusion adoption model.  Perhaps the most obvious deficiency of the diffusion model is the assumption that the nature of the externality experienced by an individual is dictated primarily by the fraction of the overall population that has adopted the service, rather than capturing the influence of adoption on each user through an influence graph, e.g., \cite{KemKle2003}, or some related model.\footnote{E.g., consider a ``model'' where the externality impact on user $i \in [N]$ may takes the form $e_i(S(t))$, where $S(t) \subset [N]$ is the subset of users having adopted the service at time $t$, and $e_i : 2^{[N]} \to \Rbb$ describes the influence of each possible subset on user $i$'s decision \cite{Web2016}.}  The diffusion adoption model, in all its variants, has long held academic interest because of its effective parsimony -- it is one of the simplest tractable models capable of capturing most (but surely not all) of the important aspects of ``real-world'' observed adoption dynamics.  

Finally, we mention the practical difficulty in obtaining useful data from real technology deployments to ``validate'' the model we have described.  One of the inherent difficulties in model validation in this context is lack of a control and variable.  A control would be a deployment of a service without any subsidy, and a variable would be a deployment of a service with a subsidy.  A scientifically rigorous model validation would be best accomplished by measuring the adoption level with and without a subsidy in identical environments.  Unfortunately this seems almost impossible in almost any practical setting due to the considerable variability across deployments of services in different target populations.  

\section{Conclusion}
\label{sec:conc}

The two-target adoption subsidy (TTAS) allows a cost- and delay- sensitive service provider to efficiently subsidize customer cost, so as to leverage the impact of the externality.  A provider focused on delay may consider the special case of the quickest adoption subsidy (QAS), or, if a finite number of subsidy adjustments are desired, the approximate QAS (AQAS).  Knowledge of equilibria and their stability is essential to properly set the target adoption level for terminating the subsidy.  This requires knowledge of the  population's affinity distribution, which may be estimated using a trial, as described in the case study.  The joint impact of the externality, service cost, and affinity distribution on adoption delay and aggregate cost motivates careful subsidy design.  

\begin{acks}
The author acknowledges Roch Gu\'{e}rin for his helpful assistance on preliminary versions of this work, as well as helpful feedback from the Associate Editor and the three anonymous reviewers.
\end{acks}

\bibliographystyle{ACM-Reference-Format-Journals}
\bibliography{ACM-ToIT-2016}

\appendix

\clearpage
\section{Nonlinear externalities}
\label{app:nlext}

In this appendix we briefly consider nonlinear externalities.  Our presentation in this section is brief and informal, with the simple intent of providing some intuition for how more general externalities will affect the previous results.  We replace the $e x$ term in $\nu(x,u)$ with $e \kappa(x)$, for $\kappa$ positive and increasing, yielding $\nu_{\kappa}(x,u) \equiv (c-u) - e {\kappa}(x)$.  The adoption dynamics hence become $\dot{x} = \gamma f_{\kappa}(x,u)$, where $f_{\kappa}(x,u) = \bar{F}_A(\nu_{\kappa}(x,u)) - x$ replaces $f(x,u)$ in \eqref{eq:adoptdyn}.  Finally, $\Xmc_{\kappa}$ denotes the set of equilibria, $\tu_{\kappa}$ denotes the subsidy completion time, and $J_{\kappa}[u]$ denotes the aggregate cost to the provider.  In fact, we consider ${\kappa}(x;\alpha) \equiv x^{\alpha}$, for $\alpha > 0$, and will then use the subscript $\alpha$ instead of ${\kappa}$.  Observe ${\kappa}(x;\alpha) > x$ ($<x$) for $\alpha <1$ ($>1$) indicates a super- (sub-) linear externality.

\begin{figure}[ht]
\centering
\includegraphics[width=0.48\columnwidth]{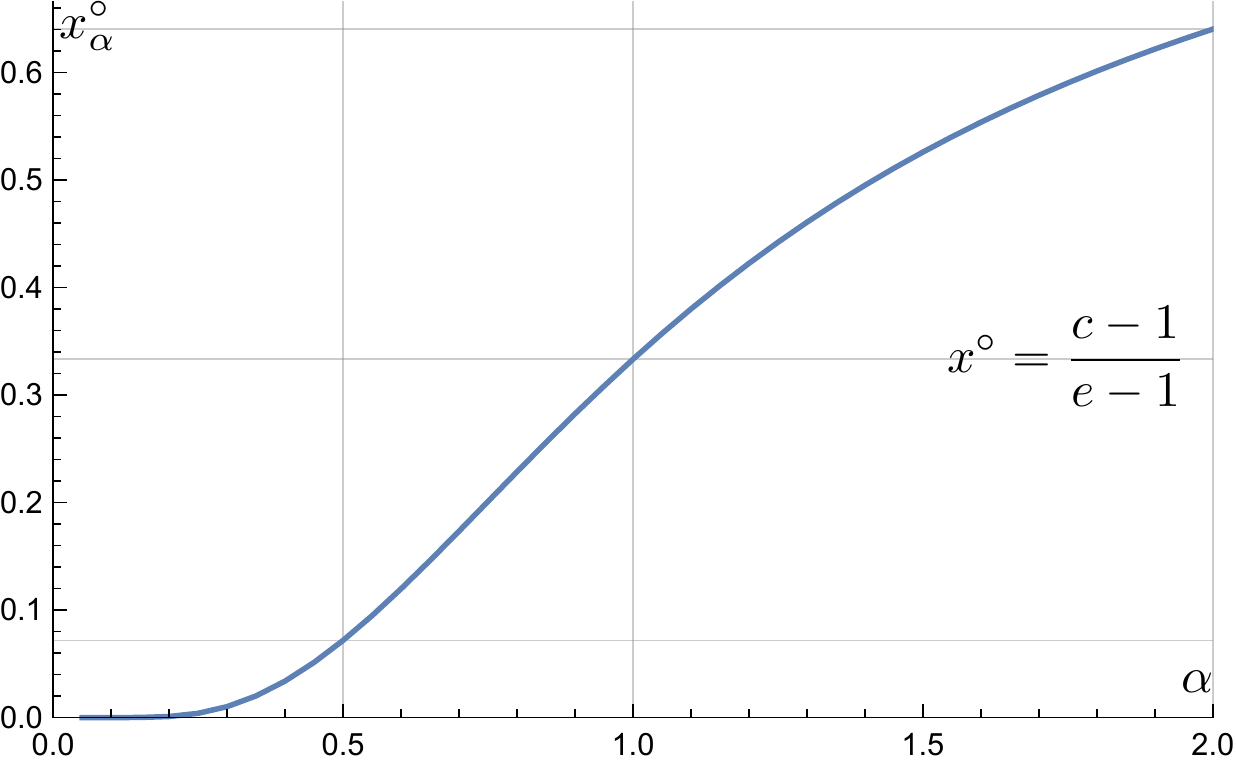}
\includegraphics[width=0.48\columnwidth]{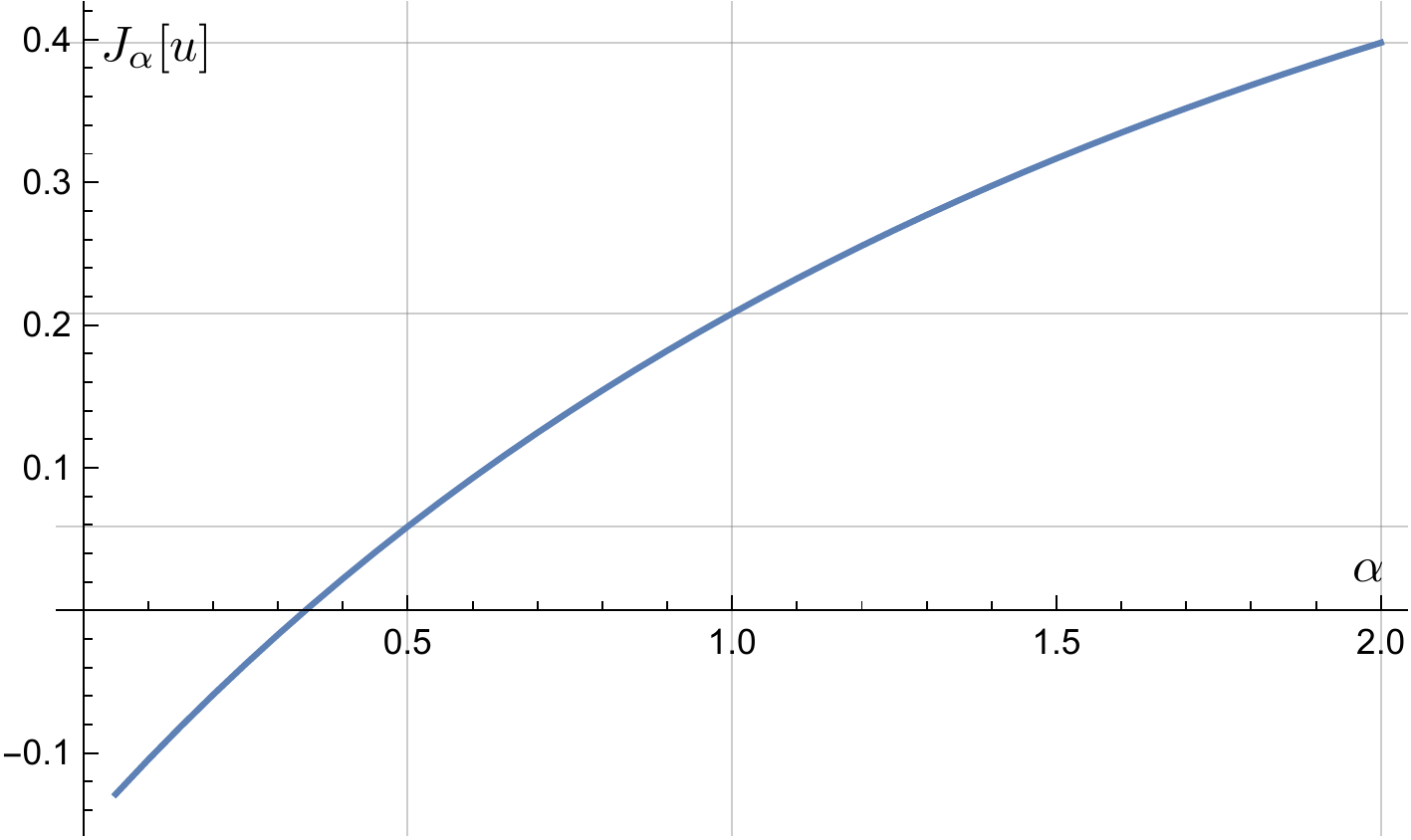}
\caption{Left: impact of nonlinear externality ${\kappa}(x;\alpha) = x^{\alpha}$ on boundary equilibrium $x_{\alpha}^{\circ}$ (recall \eqref{eq:xcircc}): superlinear externalities ($\alpha < 1$) decrease $x^{\circ}_{\alpha}$ relative to $\alpha=1$, while sublinear externalities increase it.  Gridlines show $x_{1/2}^{\circ} \approx 0.07$ and $x_2^{\circ} \approx 0.64$. Right: impact of nonlinear externality ${\kappa}(x;\alpha) = x^{\alpha}$ on the normalized aggregate cost, $\frac{\gamma}{e} J_{\alpha}[u]$, from \eqref{eq:twotarcostprfalpha}: superlinear externalities ($\alpha < 1$) decrease $J_{\alpha}[u]$ relative to $\alpha=1$, while sublinear externalities increase it.  Gridlines show cost of approximately $0.059$, $0.21$, and $0.40$ at $\alpha = 1/2, 1, 2$, respectively.}
\label{fig:nlext1}
\end{figure}

Recall from \defref{equil} that $x_{\kappa} \in \Xmc_{\kappa}$ if $f_{\kappa}(x_{\kappa},0) = 0$, meaning $\bar{F}_A(\nu_{\kappa}(x_{\kappa},0)) = x_{\kappa}$, and thus $e {\kappa}(x_{\kappa}) + \bar{F}_A^{-1}(x_{\kappa}) = c$.  The latter expression makes clear that superlinear (sublinear) $g$ decreases (increases) the equilibrium $x_{\kappa}$ relative to $x$, the equilibrium under a linear externality.  For example, for uniform affinities, $A \sim \mathrm{Uni}[0,1]$ and ${\kappa} = {\kappa}(x;\alpha)$ above, denote by $x_{\alpha}^{\circ}$ (recall \eqref{eq:xcircc}) the equilibrium at the boundary between the $0/1$ convergence regions, shown in \figref{nlext1} (left).  

Suitable adjustments in the proof of \prpref{neccondext} on cost-extremal subsidies, namely,
\begin{equation}
\dd{x}g_{\kappa}(\dot{x},x,u) = -\gamma (e f_A(\nu_{\kappa}(x,u)) \kappa'(x) - 1)
\end{equation}
yields the necessary condition for extremality as $u_{\kappa} = e x \kappa'(x) - \gamma \lambda$ instead of \eqref{eq:neccondext}.  If $\kappa'(x) > 1$ ($<1$) then $u_{\kappa}$ is larger (smaller) than the $u$ for a linear externality.  

Finally, making suitable adjustments to \defref{twotargetadsub}, namely, $\nu_{\kappa}(x,u) = \rho$ requires $u_{\kappa} = c - \rho - e \kappa(x)$.  Combining with the necessary extremality condition for $u_{\kappa}$ has solution:
\begin{equation}
(\alpha+1) e x^{\alpha} = c - \rho + \gamma \lambda, ~ 
(\alpha+1) u = \alpha (c-\rho) - \gamma \lambda,
\end{equation}
for which \eqref{eq:twotarnecprf} is now a special case of the above for $\alpha = 1$.  The corresponding normalized aggregate cost in \eqref{eq:twotarcostprf1} is now
\begin{equation}
\label{eq:twotarcostprfalpha}
\frac{\gamma}{e} J_{\kappa}[u] = \int_{\xl}^{\xu} \frac{x(\xi(\chi) - \kappa(x))}{\chi -x} \drm x,
\end{equation}
which is shown in \figref{nlext1} (right) as a function of $\alpha$.  In summary, these results accord with intuition, namely, that super- (sub-) linear externalities lower (raise) stability thresholds and decrease (increase) the cost of a subsidy, relative to linear externality, respectively.

\section{Proofs}
\label{app:proofs}

\prpref{nosubAD2} gives the AD under uniform affinities for the cases not covered in \prpref{nosubAD}.
\begin{proposition}
\label{prp:nosubAD2}
The AD under uniform affinities for cases 1, 2, and 4 are:
\begin{enumerate}
\item[1.] If $c>1,c>e$, the AD are
\begin{equation}
\label{eq:adonecolcase1}
x_1(t|\tl,\xl) = \left\{ \begin{array}{l}
\xl \erm^{-\gamma (t-\tl)} \\
x^{\circ} + (\xl - x^{\circ}) \erm^{(e-1)  \gamma (t-\tl)} \\
\frac{c-1}{e} \erm^{-\gamma(t-\tl-\hat{T}_1(\xl))} 
\end{array} \right. 
\end{equation}
where the three subcases correspond to $i)$ $\xl < \frac{c-1}{e}$, $ii)$ $\frac{c-1}{e} \leq \xl$ and $t - \tl \leq \hat{T}_1(\xl)$, and $iii)$ $\frac{c-1}{e} \leq \xl$ and $t - \tl \geq \hat{T}_1(\xl)$, respectively.  
\item[2.] If $e<c<1$, the AD are 
\begin{equation}
x_2(t|\tl,\xl) = x^{\circ} + (\xl - x^{\circ}) \erm^{-(1-e)  \gamma (t-\tl)}.
\end{equation}
\item[4.] If $c < 1, c < e$, the AD are
\begin{equation}
\label{eq:adonecolcase4}
x_4(t|\tl,\xl) = \left\{ \begin{array}{l}
1 - (1-\xl) \erm^{-\gamma (t-\tl)} \\
x^{\circ} + (\xl - x^{\circ}) \erm^{(e-1)  \gamma (t-\tl)} \\
1 - \left(1 - \frac{c}{e}\right) \erm^{-\gamma (t-\tl-\hat{T}_0(\xl))} 
\end{array} \right. 
\end{equation}
where the three subcases correspond to $i)$ $\frac{c}{e} \leq \xl$, $ii)$ $\xl \leq \frac{c}{e}$ and $t-\tl \leq \hat{T}_0(\xl)$, and $iii)$ $\xl \leq \frac{c}{e}$ and $t-\tl \geq \hat{T}_0(\xl)$, respectively.
\end{enumerate}
\end{proposition}
\begin{proof}[of \prpref{nosubAD2}]
Separation of variables on the AD gives
\begin{equation}
\label{eq:uniaffADsolsepvar}
\frac{1}{c-w-e \bar{F}_W(w)} \drm w = \gamma \drm t.
\end{equation}
The antiderivative of the LHS is 
\begin{equation}
\label{eq:uniaffFWcases}
\int \frac{1}{c-w-e \bar{F}_W(w)} \drm w = \left\{ \begin{array}{l}
-\log(c-e-w) \\
-\frac{1}{1-e} \log(c-e-(1-e)w) \\
-\log(c-w)
\end{array} \right.,
\end{equation}
where the three subcases correspond to $i)$ $w < 0$, $ii)$ $0 \leq w < 1$, and $iii)$ $w \geq 1$, respectively.  It is necessary to consider each of the four $(e,c)$ cases separately; for conciseness we present only the analysis for case $1$ ($c > 1$ and $e > 1$), as the other cases  require a similar analysis.  Let $w_0 = c - e \xl$ be the initial value $w(\tl)$, where $\xl=x(\tl)$.  From the bottom of Fig.~\ref{fig:uniaffeqmap} it is clear there are two subcases: $i)$ $w_0 \in [1,c]$ and $ii)$ $w_0 \in [c-e,1]$.  First suppose $w_0 \in [1,c]$, corresponding to subcase $iii)$ in \eqref{eq:uniaffFWcases}, where \eqref{eq:uniaffADsolsepvar} has solution 
\begin{equation}
\label{eq:uniaffADsolpfwt}
w(t|\tl,w_0) = c - (c-w_0) \erm^{-\gamma (t-\tl)}, ~ t \geq \tl.
\end{equation}
Second, suppose $w_0 \in [c-e,1]$, for which subcase $ii)$ of \eqref{eq:uniaffFWcases} will determine the solution for $t \in [\tl,t_w]$ and then subcase $iii)$ of \eqref{eq:uniaffFWcases} will determine the solution for $t > t_w$, where $t_w$ obeys $w(t_w) = 1$. Over $t \in [\tl,t_w]$ the solution of \eqref{eq:uniaffADsolsepvar} is 
\begin{equation}
w(t)  = \frac{c-e}{1-e} - \left(\frac{c-e}{1-e} - w_0 \right)\erm^{-(1-e)\gamma(t-\tl)}, 
\end{equation}
and solving $w(t_w) = 1$ for $t_w$ gives
\begin{equation}
\label{eq:uniafftw}
t_w = \tl - \frac{1}{\gamma(1-e)} \log\left(\frac{c-1}{c-e-(1-e)w_0}\right).
\end{equation}
Finally, for $t > t_w$ the solution is again \eqref{eq:uniaffADsolpfwt}, but with initial condition $\tl = t_w$ and $w_0 = 1$.  Finally, the dynamics for $x(t)$ are recovered from the solution for $w(t)$ via $x(t) = (c-w(t))/e$, $\xl = (c-w_0)/e$, and $x^{\circ}$ in \eqref{eq:xcircc}.  
\end{proof}

\begin{figure}[!ht]
\centering
\includegraphics[width=0.75\linewidth]{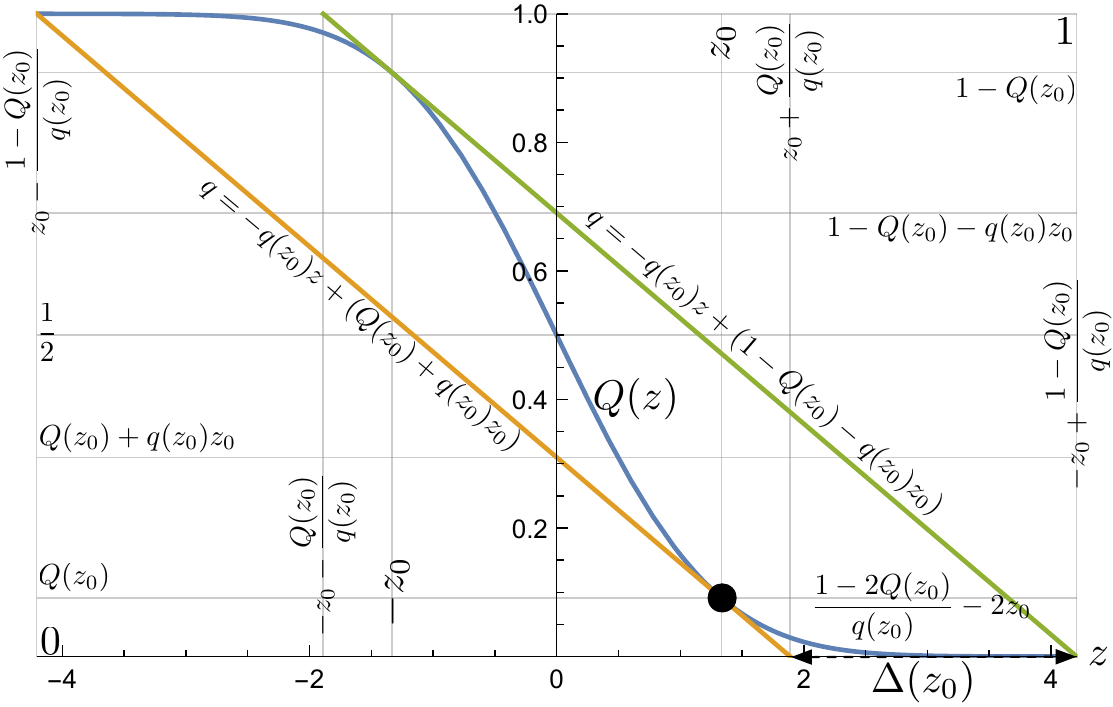}
\caption{(\secref{noraff}): Proof of \prpref{normaffnumeq} for the point $(z_0,Q(z_0))$ (black circle); vertical (horizontal) labels denote vertical (horizontal) lines, respectively.}
\label{fig:normaffeqmapproof}
\end{figure}

\begin{proof}[of \prpref{normaffnumeq}]
\label{prf:vide43vg34}
See \figref{normaffeqmapproof}. For any $z_0 > 0$ the tangent line at point $(z_0,Q(z_0))$ (black circle) to $Q(z)$ (blue curve) is $q_1(z) = -q(z_0) z + (Q(z_0) + q(z_0) z_0)$ (orange line) with slope $-q(z_0)$ and $q$-axis intercept $Q(z_0) + q(z_0) z_0$.  The line $q_2(z) = -q(z_0) z + (1-Q(z_0)-q(z_0) z_0)$ (green line) is parallel with the $q$-axis intercept $1-Q(z_0)-q(z_0) z_0$, and tangent to $Q(z)$ at $(-z_0,1-Q(z_0))$.  The horizontal distance between the lines is 
\begin{equation}
\label{eq:normaffDelta}
\Delta(z_0) \equiv \frac{1-2 Q(z_0)}{q(z_0)} - 2 z_0.
\end{equation}
Consider the set of all lines parallel to $q_1(z)$, parameterized by the horizontal displacement $d$, denoted $\Lmc(z_0) = \{ q_1(z-d), ~ d \in \Rbb\}$, and let $\Xmc(z_0,d)$ denote the set of intersections of the line $q_1(z-d)$ with the curve $Q(z)$.  From the figure:
\begin{equation}
|\Xmc(z_0,d)| = \left\{ \begin{array}{ll}
2, \; & d  \in \{0,\Delta(z_0)\} \\
1, \; & d < 0 \mbox{ or } d > \Delta(z_0) \\
3, \; & d \in (0,\Delta(z_0))
\end{array} \right.
\end{equation}
That is, the line $q_2(z)$ with displacement $\Delta(z_0)$ is the critical case of displacements with three intersections with $Q(z)$. 

Observe the adoption level interval $x \in [0,1]$ implies the reparameterized adoption level $z = c - ex \in [c-e,e]$.  From Def.~\ref{def:equil}, the set of equilibria is $\Xmc = \{ z : (c-z)/e = Q(z)\}$; in words, $\Xmc$ is the set of intersections of $Q(z)$ with the line segment on the $(z,q)$ plane of slope $-1/e$ connecting the points $(c-e,1)$ and $(c,0)$.  Observe $\Xmc \neq \emptyset$ for all $(e,c)$.  Observe the one-to-one correspondence between $e \in [\sqrt{2\pi},\infty)$ and $z_0(e)$ via $z_0(e) = q^{-1}(1/e)$, for $q^{-1}$ in \eqref{eq:normalpdfinv}, \ie for any $1/e < 1/\sqrt{2\pi}$ there exists a unique $z_0(e) > 0$ such that the line $q_1$ defined by $z_0(e)$ above has slope $-1/e$.  The line segment $(c-z)/e = (-1/e)z + (c/e)$ with $z$-axis intercept $c$ will have three intersections with $Q(z)$ iff $c$ lies between the $z$-axis intercepts of the orange and green lines in \figref{normaffeqmapproof}, \ie 
\begin{equation}
z_0(e) + \frac{Q(z_0(e))}{q(z_0(e))} < c < z_0(e) + \frac{Q(z_0(e))}{q(z_0(e))} +\Delta(z_0(e)),
\end{equation}
which simplifies to $c \in (c_l(e),c_u(e))$.  Observe \eqref{eq:normaffdelta} and \eqref{eq:normaffDelta} are related via $\delta(e) = \Delta(z_0(e))$ for $z_0(e)=q^{-1}(1/e)$.
\end{proof}

\begin{proof}[of \eqref{eq:b9f34gf34} in \secref{case}] 

{\em Time unit.} Recall from \remref{modeljust} (point (1)) in \secref{assjust} that utilities (including affinities) and costs are measured per unit time.  For example, if the unit is a week then $A$ is the inherent value (in dollars) of the service to a user per week and $c$ is the subscription cost (in dollars) over a week.  The time unit is not the subscription commitment duration nor the billing period, however, as users are empowered to adjust their subscription decisions on the finer time scale of the time slot, and moreover users are assumed to be billed per time slot, as discussed below.  Although the time unit is arbitrary, it naturally determines the range of values, i.e., the ``effective support size'', of both the affinity distribution $F_A$ and the value of $c$.  In the case of uniformly distributed affinities, $A \sim \mathrm{Uni}[\underline{a},\bar{a}]$, this effective support size may naturally be defined as $\bar{a}-\underline{a}$, while in the case of normally distributed affinities, $A \sim \Nmc(\mu,\sigma)$, the effective support size is naturally defined as $\sigma$.  As discussed in \secref{aff}, in both these cases there is no loss in generality in {\em scaling} the model such that the effective support size is unity: for uniformly distributed affinities one scales from $(\underline{a},\bar{a},c',e')$ to $(0,1,c,e)$, and for normally distributed affinities one scales from $(\mu,\sigma,c',e')$ to $(0,1,c,e)$.  In both these cases the subsequent analysis identified $c=1$ as a critical cost per unit time (with respect to the nature of the equilibria) when the model was scaled to have a unit effective support size.  This model scaling corresponds to selecting a time unit.  

{\em Discrete-time finite population AD.} Let time be slotted, with time (slots) indexed by $t \in \Nbb$ and of duration $h$ time units per time slot, for $h<1$.  The notion of a time slot is intended to capture the smallest period of time over which users make service subscription decisions.  That is, we suppose utilities and costs are measured per unit time, subscription decisions are made per time slot, and there are multiple ($1/h$) time slots per unit time.  We will show that these discrete-time dynamics approach the continuous-time dynamics \eqref{eq:9fd9034f34g} as $h \downarrow 0$.  Continuing the example where the time unit is a week, if the time slot duration is set at one day then $h=1/7$.  

Consider a (fixed) finite population of $N \in \Nbb$ potential service subscribers, indexed by $i \in [N]$, with heterogeneous random affinities (per time unit) given by the iid sequence $(A_i, i \in [N])$, for $A_i \sim F_A$ and $i \in [N]$.  Let $x_t \equiv \frac{1}{N} \sum_{i=1}^N b_{i,t} \in [0,1]$ denote the fraction of the population that has chosen to adopt the service for time slot $t$, with $x_0$ the (given) initial adoption level, and $b_{i,t} \in \{0,1\}$ the time slot $t$ adoption decision of user $i$.  Thus $(x_t, t \in \Zbb_+)$ is the discrete-time adoption process.  The following steps are assumed to happen at the instant between the end of slot $t$ and the beginning of slot $t+1$:
\begin{itemize}
\itemsep=-2pt
\item The service controller advertises to the population the value of $x_t$ for the (just-ended) time slot $t$, then computes (on the basis of $x_t$) and advertises to the population a subsidy level $u_{t+1}$ (per time unit) for the upcoming time slot.
\item Each potential user $i \in [N]$ decides whether or not to evaluate her current subscription decision by flipping a coin $\sigma_{i,t+1} \sim \mathrm{Ber}(h \gamma) \in \{0,1\}$, for $\gamma \in (0,1/h)$, with $(\sigma_{i,t+1},i\in [N],t\in \Nbb)$ iid in both $(i,t)$.  Observe each user waits on average $1/(h\gamma)$ time slots before reassessing her subscription decision, giving the discrete-time AD a time scale of $1/\gamma$ time units.  For example, if $\gamma=1/4$ then each user waits on average $28$ days (time slots) or $4$ weeks (time units) between subscription reassessments.
\item Each potential user's estimated utility / perceived value per time unit of the service for the upcoming time slot $t+1$ at this point in time is assumed to equal:
\begin{equation}
V_i(x_t,u_{t+1}) \equiv A_i + e x_t - (c - u_{t+1}), ~ i \in [N].
\end{equation}
Note the utility estimates are made on the basis of $x_t$ and $u_{t+1}$, as is natural, and that $e$ has units of dollars per time unit.
\item The random user subset $\Imc_{t+1} = \{ i \in [N] : \sigma_{i,t+1} = 1 \}$ each make subscription decisions $b_{i,t+1} = \mathbf{1}(V_i(x_t,u_{t+1}) > 0)$ for the (about to begin) time slot $t+1$.
\item The remaining users $[N] \setminus \Imc_{t+1}$ carry their time $t$ subscription decisions forward to time $t+1$, i.e., $b_{i,t+1} = b_{i,t}$.  
\item The service controller bills all subscribers for the upcoming time slot, i.e., those with $b_{i,t+1} = 1$, the amount $h(c-u_{t+1})$ for the upcoming time slot $t+1$.
\end{itemize}

{\em Analysis.} Under these assumptions the discrete-time AD is given by \eqref{eq:b9f34gf34}.  As $i)$ $u_{t+1}$ is a function of $x_t$, and $ii)$ $x_t$ is in turn a function of $b_t \equiv (b_{i,t}, i \in [N])$, it follows that:
\begin{equation}
\Ebb[x_{t+1}|b(t)] = h \gamma \frac{1}{N} \sum_{i \in [N]} \mathbf{1}(V_i(x_t,u_{t+1}) > 0) + (1-h\gamma) x_t.
\end{equation}
As $(A_i, i \in [N])$ are iid, the Law of Large Numbers ensures 
\begin{equation}
\lim_{N \uparrow \infty} \frac{1}{N} \sum_{i \in [N]} \mathbf{1}(V_i(x_t,u_{t+1}) > 0) = \Ebb[\mathbf{1}(V(x_t,u_{t+1}) > 0)] = \Pbb(V(x_t,u_{t+1}) > 0).
\end{equation}
For large but finite $N$ this probability is an approximation of the population average:
\begin{equation}
\Ebb[x_{t+1}|b_t] \approx h \gamma \Pbb(V(x_t,u_{t+1}) > 0) + (1-h\gamma) x_t,
\end{equation}
with the understanding that the approximation becomes exact in the limit as $N \uparrow \infty$.  The above expression may be rearranged as
\begin{equation}
\Ebb \left[ \left. \frac{x_{t+1}-x_t}{h} \right|b_t \right] \approx \gamma \left( \Pbb(V(x_t,u_{t+1}) > 0) - x_t \right).
\end{equation}
Observing $\lim_{h \downarrow 0} (x_{t+1}-x_t)/h = \dot{x}(t)$ we recover the continuous-time AD \eqref{eq:9fd9034f34g} for large $N$.  For small but finite $h$ we have an accurate approximation of \eqref{eq:9fd9034f34g}.
\end{proof}

\end{document}